\documentclass[lettersize,journal]{IEEEtran}
\usepackage{cite}
\usepackage{cite,balance}
\usepackage{amsmath,amssymb,amsfonts,nicematrix}
\usepackage{bbm}
\usepackage{xcolor}
\usepackage{algorithm2e}
\usepackage{algcompatible}
\usepackage{algpseudocode}
\def\BibTeX{{\rm B\kern-.05em{\sc i\kern-.025em b}\kern-.08em
    T\kern-.1667em\lower.7ex\hbox{E}\kern-.125emX}}
\usepackage{nicematrix,tikz}
\usepackage{graphicx}
\usepackage{textcomp}
\usepackage{array,booktabs,makecell}
\usepackage[font=footnotesize]{caption}
\usepackage{float}
\usepackage{tikz}
\usepackage{cite,balance}
\usetikzlibrary{shapes.geometric, arrows}
\usepackage{pgfplots}
\usepackage{caption}
\usepackage{mathtools}
\usepackage{subcaption}
\usepackage{comment}
\usepackage{enumitem}
\usepackage{stackengine}

\newcommand\numberthis{\addtocounter{equation}{1}\tag{\theequation}}
\usepackage{xcolor}
\definecolor{OliveGreen}{rgb}{0,0.6,0}
\def\BibTeX{{\rm B\keri-.05em{\sc i\keri-.025em b}\keri-.08em
    T\keri-.1667em\lower.7ex\hbox{E}\keri-.125emX}}
\usepackage{multirow}

\usepackage{amsthm}
\newtheorem{remark}{Remark}

\newtheorem{definition}{Definition}
\newtheorem{theorem}{Theorem}

\newtheorem{lemma}{Lemma}

\newtheorem{design}{Design}
\newtheorem{subcase}{Subcase}
\allowdisplaybreaks
\usepackage{mathtools}
\newtheorem{assumption}{Assumption}
\usepackage{pifont}

\SetKwComment{Comment}{/* }{ */}
\RestyleAlgo{ruled}
\newcommand{\defeq}{\vcentcolon=}

\setlist[enumerate]{wide=0pt, leftmargin=15pt, labelwidth=15pt, align=left}
\usepackage{tabstackengine}

\definecolor{bleudefrance}{rgb}{0.19, 0.55, 0.91}
\definecolor{dollarbill}{rgb}{0.52, 0.73, 0.4}
\definecolor{fuchsiapink}{rgb}{1.0, 0.47, 1.0}
\definecolor{mikadoyellow}{rgb}{1.0, 0.77, 0.05}
\definecolor{tangerine}{rgb}{0.95, 0.52, 0.0}
\definecolor{slateblue}{rgb}{0.42, 0.35, 0.8}
\definecolor{meatbrown}{rgb}{0.9, 0.72, 0.23}
\definecolor{lavenderpink}{rgb}{0.98, 0.68, 0.82}
\definecolor{lavenderrose}{rgb}{0.98, 0.63, 0.89}
\definecolor{lavendermagenta}{rgb}{0.93, 0.51, 0.93}
\definecolor{jonquil}{rgb}{0.98, 0.85, 0.37}
\definecolor{heliotrope}{rgb}{0.87, 0.45, 1.0}
\definecolor{electriclavender}{rgb}{0.96, 0.73, 1.0}
\definecolor{spirodiscoball}{rgb}{0.06, 0.75, 0.99}
\definecolor{azure(colorwheel)}{rgb}{0.0, 0.5, 1.0}
\definecolor{amethyst}{rgb}{0.6, 0.4, 0.8}
\definecolor{ao(english)}{rgb}{0.0, 0.5, 0.0}
\definecolor{darkcyan}{rgb}{0.0, 0.55, 0.55}

\usepackage{bm}
\usetikzlibrary{arrows.meta}
\tikzset{every picture/.style={line width=0.6pt}}
\usetikzlibrary{calc,positioning}
\usetikzlibrary{arrows.meta,
                backgrounds,
                chains,
                fit,
                quotes}

\newcommand*\circled[1]{\tikz[baseline=(char.base)]{
    \node[shape=circle,draw,inner sep=0.5pt] (char) {#1};}}

\begin{document}
\title{Cooperative Gradient Coding}
\author{Shudi Weng,~\IEEEmembership{Graduate Student Member,~IEEE,}
Chao Ren,~\IEEEmembership{Member,~IEEE,}\\
Ming Xiao,~\IEEEmembership{Senior Member,~IEEE,}
and Mikael Skoglund,~\IEEEmembership{Fellow,~IEEE}\vspace{-1em}
\thanks{Shudi Weng, Ming Xiao, Chao Ren, and Mikael Skoglund are with the School of Electrical Engineering and Computer Science (EECS), KTH Royal Institute of Technology, 11428 Stockholm, Sweden, Email: \{shudiw, mingx, chaor, skoglund\}@kth.se. Corresponding author: Shudi Weng.}}

\markboth{Journal of \LaTeX\ Class Files,~Vol.~18, No.~9, September~2020}%
{How to Use the IEEEtran \LaTeX \ Templates}

\maketitle

\begin{abstract}
This work studies gradient coding (GC) in the context of distributed training problems with unreliable communication. We propose cooperative GC (CoGC), a novel gradient-sharing-based GC framework that leverages cooperative communication among clients. This approach ultimately eliminates the need for dataset replication, making it both communication- and computation-efficient and suitable for federated learning (FL). {By employing the standard GC decoding mechanism, CoGC yields strictly binary outcomes: either the global model is exactly recovered, or the decoding fails entirely, with no intermediate results. This characteristic ensures the optimality of the training and demonstrates strong resilience to client-to-server communication failures when the communication channels among clients are in good condition. However, it may also result in communication inefficiency and hinder convergence due to its lack of flexibility, especially when communication channels among clients are in poor condition.} To overcome this limitation and further harness the potential of GC matrices, we propose a complementary decoding mechanism, termed GC$^+$, which leverages information that would otherwise be discarded during GC decoding failures. This approach significantly improves system reliability under unreliable communication, as the full recovery of the global model typically dominates in GC$^+$.
To conclude, this work establishes solid theoretical frameworks for both CoGC and GC$^+$. We provide complete outage analyses for each decoding mechanism, along with a rigorous investigation of how outages affect the structure and performance of GC matrices. Building on these analyses, we derive convergence bounds for both decoding mechanisms. Finally, the effectiveness of CoGC and GC$^+$ is validated through extensive simulations.

\end{abstract}


\section{Introduction}\label{Sec:Intro}
\IEEEPARstart{D}{istributed} learning (DL) is a paradigm shift in accelerating machine learning (ML) model training, driven by the recent developments of computing resources and the vast amount of data generated in the Internet-of-Things (IoT) networks\cite{9562559,qian2022distributed}. DL leverages the edge computational power to solve a global minimization problem, thereby alleviating the computational bottleneck in classic centralized ML approaches. However, the dataset transmission in DL requires excessive communication overhead. To address this, federated learning (FL) has emerged as a promising method by restricting raw dataset replication despite keeping the same training process as DL. This constraint not only reduces the communication overhead but also enhances privacy preservation to some extent.

Extensive studies on DL and FL are carried out under the assumption of perfect communication \cite{pmlr-v70-wang17f, 9855231}. However, in real-world scenarios, various physical factors, (e.g., fading, shadowing, and network congestion, etc.), can cause intermittent connectivity between clients and the parameter server (PS). The unpredictable participation of clients can lead to a strictly suboptimal training process. For this reason, various efforts have been made to enhance the robustness of DL and FL under unreliable communication. 
Namely, Zheng et al., \cite{zheng2023federated} design an unbiased aggregation rule at the PS by taking the heterogeneous connectivity into account. Wang et al., \cite{wang2021quantized,chu2022federated} propose resource allocation strategies to balance network asymmetry. Saha et al.,  \cite{saha2022colrel,yemini2023robust} introduce collaborative relaying to improve robustness against stragglers by dynamically adjusting collaboration weights adaptive to network conditions. 
{These works \cite{yu2019distributed,wang2021quantized,zheng2023federated,saha2022colrel,yemini2023robust} have shown excellent effectiveness in handling stragglers. However, they heavily rely on precise prior knowledge of channel statistics across the entire network, limiting their practical feasibility in complex and dynamic environments.}

The problem of straggler mitigation in DL has also attracted significant interest from the coding community. {One of the most well-known contributions is gradient coding (GC), originally proposed by Tandon et al. \cite{tandon2017gradient}, which introduces cyclic gradient codes to recover the exact sum of individual gradients in the presence of stragglers.} Subsequently, a series of follow-up works have extended GC to better accommodate various practical scenarios and constraints. To further improve the straggler tolerance of GC, \cite{krishnan2021sequential} proposes sequential GC that enables recovery of the exact sum not only within a single snapshot, but also across the temporal sequences. To adapt diverse computational capabilities among clients, {\cite{wang2019heterogeneity}}, \cite{wang2021heterogeneity} proposes a heterogeneity-aware GC scheme where datasets are no longer uniformly replicated in equal amounts among clients. Following this idea, \cite{jahani2021optimal} further investigates the optimal communication-computation trade-off of heterogeneous GC in the presence of malicious parties. Another line of work focuses on improving communication efficiency in GC. Namely, Ye et al., \cite{ye2018communication} develop a communication-efficient GC scheme by encoding gradient vectors across dimensions, and establish a three-dimensional trade-off among straggler tolerance, computational load, and communication costs. Further advancements in decoding complexity and numerical stability are achieved by \cite{kadhe2020communication}, maintaining the same trade-offs while improving coding efficiency. To minimize communication costs in dynamic straggler presence, Cao et al., \cite{9609019} dynamically adjust the transmitted partial sums in response to the observed straggler patterns. To alleviate the computational load on individual clients, \cite{tang2024design} designs hierarchical GC (HGC), which involves more edge devices to compute on each replicated dataset and offloads the computation of partial sums to intermediate relay nodes. Further computation-communication trade-off is explored in \cite{gholami2025optimal}. 
These coding strategies \cite{tandon2017gradient,ye2018communication,kadhe2020communication,9609019,wang2021heterogeneity,krishnan2021sequential,jahani2021optimal} have demonstrated success in handling stragglers. {However, they adopt dataset-replication-based GC framework, limiting their applicability in several aspects. First, dataset transmission incurs excessive transmission overhead and energy consumption. Second, replicating raw datasets poses serious privacy concerns. Third, they are incompatible with FL due to the dataset-sharing constraints.}

\begin{figure}[t]
\centering
\scalebox{0.85}{\begin{tikzpicture}[
    client/.style={draw, minimum width=1.5cm, minimum height=0.8cm},
    arrow/.style={>=stealth'},
    PS/.style={draw,minimum height=0.8cm,text width=4cm, text centered}
]

\node[PS] (top) at (0,2) {Any two: $\boldsymbol{g}_1+\boldsymbol{g}_2+\boldsymbol{g}_3$};

\node[client] (b1a) at (-2.5,0) {$\mathcal{D}_1$};
\node[client] (b1b) [below=-0.025cm of b1a] {$\mathcal{D}_2$};

\node[client] (b2a) at (0,0) {$\mathcal{D}_2$};
\node[client] (b2b) [below=-0.025cm of b2a] {$\mathcal{D}_3$};

\node[client] (b3a) at (2.5,0) {$\mathcal{D}_3$};
\node[client] (b3b) [below=-0.025cm of b3a] {$\mathcal{D}_1$};

\draw[->] (b1a.north) -- ([xshift=-0.5cm]top.south) node[above, pos=0.3, xshift=-0.4cm] {$\frac{1}{2}\boldsymbol{g}_1+\boldsymbol{g}_2$};
\draw[->] (b2a.north) -- (top.south) node[above right, pos=0.3] {$\boldsymbol{g}_2-\boldsymbol{g}_3$};
\draw[->] (b3a.north) -- ([xshift=0.5cm]top.south) node[above, pos=0.3, xshift=0.6cm] {$\frac{1}{2}\boldsymbol{g}_1+\boldsymbol{g}_3$};

\end{tikzpicture}}
\vspace{2mm}
\caption{{Dataset-replication-based GC with $M=3$ and $s=1$ \cite{tandon2017gradient}. Each client trains multiple datasets and transmits the partial sums.} } 
\label{fig:tandon_framework}
\vspace{-0.6cm}
\end{figure}
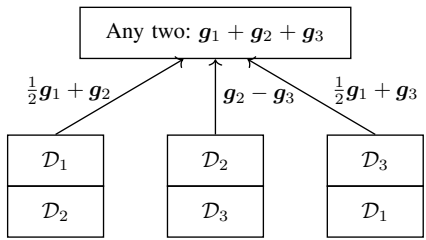

There is a pressing need to develop GC methods tailored to FL settings. In this context, Schlegel et al. \cite{schlegel2023codedpaddedfl} propose an approach that integrates dataset-replication-based GC into coded FL, enabling clients to share coded datasets rather than raw datasets. While this method improves privacy protection, it still imposes substantial communication overhead, as the transmission of redundant coded datasets requires significant power and time. {To overcome this limitation and to truly enable GC in FL, our prior work \cite{weng2024supplementaryfilecooperativegradient}, introduces a gradient-sharing-based GC framework, cooperative gradient coding (CoGC), which is implemented by gradient sharing among clients.} The CoGC is suitable for both FL and DL, it significantly reduced communication overhead, by ultimately eliminating the need for raw data sharing. Moreover, it significantly reduced computational load, since each client only processes its own single dataset in CoGC.

Another critical challenge in GC is the decoding efficiency. The existing GC regimes \cite{tandon2017gradient,ye2018communication,kadhe2020communication,9609019,wang2021heterogeneity,krishnan2021sequential,jahani2021optimal,schlegel2023codedpaddedfl,weng2024supplementaryfilecooperativegradient} are founded on the standard GC decoding mechanism, where the PS can either compute the exact sum or nothing. This binary recovery of the exact sum in the existing GC mechanism offers threefold advantages: $\textit{(i)}$ the PS can only learn about the partial sums, preventing the PS from knowing any individuals and thus enhancing privacy; $\textit{(ii)}$ the optimality of the learning algorithms is ensured by the exact recovery; and $\textit{(iii)}$ the straggler tolerance is achieved through the coding structure, eliminating the need for prior connectivity information. Yet this binary characteristic can also waste computation and communication resources, even hindering convergence in some cases. To illustrate, the standard GC decoding mechanism completely fails to extract any useful information about the global model if the number of stragglers exceeds the predefined threshold, owing to the neglect of the valuable information in the incomplete partial sums. In such a design, a high overall outage probability can lead to a considerable number of communication attempts before achieving successful recovery of the global model. Even worse, it can exacerbate divergence in local models, hindering convergence and leading to wasted computational resources. This limitation underscores the inflexibility of the standard GC decoding mechanism.

 
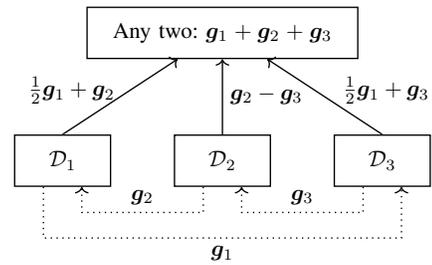
\begin{figure}[t]
\centering
\scalebox{0.85}{\begin{tikzpicture}[
    client/.style={draw, minimum width=1.5cm, minimum height=0.8cm},
    arrow/.style={>=stealth'},
    PS/.style={draw,minimum height=0.8cm,text width=4cm, text centered}
]

\node[PS] (top) at (0,2) {Any two: $\boldsymbol{g}_1+\boldsymbol{g}_2+\boldsymbol{g}_3$};

\node[client] (c1) at (-2.5,0) {$\mathcal{D}_1$};
\node[client] (c2) at (0,0) {$\mathcal{D}_2$};
\node[client] (c3) at (2.5,0) {$\mathcal{D}_3$};

\draw[->] (c1.north) -- ([xshift=-0.7cm]top.south)node[above, pos=0.3, xshift=-0.4cm] {$\frac{1}{2}\boldsymbol{g}_1+\boldsymbol{g}_2$};
\draw[->] (c2.north) -- (top.south)node[above right, pos=0.3] {$\boldsymbol{g}_2-\boldsymbol{g}_3$};
\draw[->] (c3.north) -- ([xshift=0.7cm]top.south) node[above, pos=0.3, xshift=0.6cm] {$\frac{1}{2}\boldsymbol{g}_1+\boldsymbol{g}_3$};

\draw[dotted, ->] 
    ([xshift=-0.3cm]c1.south) -- ++(0,-0.8) -- ++(5.6,0)node[below, pos=0.5, xshift=-0cm] {$\boldsymbol{g}_1$} -- ++(0,0.8);
    
\draw[dotted, ->] 
    ([xshift=-0.3cm]c2.south) -- ++(0,-0.4) -- ++(-1.9,0) node[above, pos=0.5, xshift=-0cm] {$\boldsymbol{g}_2$} -- ++(0,0.4);

\draw[dotted, ->] 
    ([xshift=-0.3cm]c3.south) -- ++(0,-0.4) -- ++(-1.9,0) node[above, pos=0.5, xshift=-0cm] {$\boldsymbol{g}_3$} -- ++(0,0.4); 

\end{tikzpicture}}
\vspace{1mm}
\vspace{-2pt}
\caption{{Our proposed gradient-sharing-based CoGC. Each client trains one dataset, and shares its trained model with certain neighbors. }} 
\label{fig:CoGC_framework}
\vspace{-0.6cm}
\end{figure}

\subsection{Contributions}
This work seeks to bridge the knowledge gaps in the existing literature.
{We propose a novel GC framework, CoGC, that incorporates cyclic GC codes into cooperative networks.} The CoGC ultimately eliminates the need for dataset replication, making it computation-communication-efficient and applicable to FL. Moreover, it naturally tackles random stragglers and heterogeneous networks without the need for prior information. 
{The existing theoretical analysis of GC problems primarily focuses on trade-offs among computational load, communication costs, and straggler tolerance, little research explores how the GC design impacts overall system performance from a holistic perspective.} Furthermore, to address the inflexibility issue of the binary recovery, this work proposes a complementary decoding mechanism for GC that leverages the information waste in incomplete partial sums when the standard GC decoding fails, enabling the retrieval of individual local model updates to enhance the robustness against stragglers. Compared to the short, preliminary conference version \cite{weng2024supplementaryfilecooperativegradient}, the main contributions are summarized as follows:  
\vspace{0.5em}\\
\noindent(\textbf{C1}) To advance the theoretical understanding of the GC problem within practical learning scenarios, this work links GC design and learning metrics. Specifically, we establish the essential principles for effective GC application, and conduct a general outage analysis of CoGC, which statistically quantifies the system reliability provided by the standard GC decoding mechanism. Furthermore, we quantify the convergence bound of 99.86\% given a large number of training rounds $T$, which is relaxed compared to \cite{weng2024supplementaryfilecooperativegradient}. 
\vspace{0.5em}\\
\noindent(\textbf{C2}) To ensure a complete analysis of the standard GC decoding mechanism, we evaluate its secure aggregation properties and quantify the information leakage of individual local models in partial sums. To the best of our knowledge, no prior work considers the secure aggregation feature of GC. 
\vspace{0.5em}\\
(\textbf{C3}) To further reduce communication costs while ensuring system performance, this work formulates a cost-efficient GC design problem, linking the cyclic GC code design to the target performance. The proposed strategy can significantly reduce communication and is verified through simulations.
\vspace{0.5em}\\
(\textbf{C4}) To overcome the limitation of the standard GC decoding mechanism, this work proposes a complementary decoding mechanism, GC$^+$, which leverages information waste in GC failure scenarios to enable the retrieval of any possible subset of individual local models.
The necessity of enhancing GC is discussed throughout the paper, and a complete theoretical analysis is provided for GC$^+$. Specifically, the robustness provided by GC$^+$ and the perturbed GC codes is evaluated through rank analysis, followed by convergence analysis. The GC$^+$ allows for more decoding flexibility, and its performance gains is validated through simulations.
\subsection{Paper Organization}
The remainder of this work is organized as follows. Section  \ref{sec: System Model} presents our system model and the standard GC decoding mechanism. Next, the proposed CoGC is introduced in Section \ref{sec: CoGC}, followed by the cost-efficient design of GC codes in Section \ref{sec: Cost-eff GC}. The proposed complementary decoding mechanism, GC$^+$, is introduced in Section \ref{sec: GC$^+$}. In Section \ref{sec: simulation}, the effectiveness of CoGC and GC$^+$ is evaluated via numerical simulations. Finally, this work is concluded in Section \ref{sec: conclusion}. Supporting proofs are provided in the appendices.

\section{System Model and Preliminaries}\label{sec: System Model}
\subsection{Distributed Training} 
Despite DL and FL form datasets in different ways, they share the same training process. W.L.O.G., let us consider a classic distributed setup consisting of a central PS and $M$ clients. The system aims to collaboratively solve the following empirical risk minimization (ERM) problem:
\begin{align*}
    \min_{\boldsymbol{g}\in \mathbb{R}^D}\left\{F(\boldsymbol{g})\defeq \sum_{m=1}^{M}\omega_mF_m(\boldsymbol{g}, \mathcal{D}_m)\right\},
    \numberthis
    \label{eq:goal_FL}
\end{align*}
by defining the global objective function (GOF) $F(\cdot)$ as a weighted sum of local objective functions (LOFs) $\{F_m: \mathbb{R}^D\times \mathcal{D}_m\rightarrow \mathbb{R}\}$ at clients $m\in\{1, \cdots, M\}$, where $F_m(\boldsymbol{g})=\frac{1}{\lvert \mathcal{D}_m \rvert}\sum_{\xi\in \mathcal{D}_m}\mathcal{L}(\boldsymbol{g},\xi)$ is the averaged loss $\mathcal{L}(\cdot)$ evaluated on the entire local dataset $\mathcal{D}_m$. Clients can have different levels of importance, reflected in the learning weight $\omega_m$. The commonly used method to determine this weight is by the relative size of the local datasets, i.e., $\omega_m=\frac{\lvert \mathcal{D}_m \rvert}{\sum_{m=1}^M \lvert \mathcal{D}_m \rvert }$. For clarity and W.L.O.G., in this article, all clients are assumed to be equally important, i.e., $\omega_m=\frac{1}{M}$.


After receiving the latest global model, each client performs $I$-step consecutive local training empowered by its local computational capabilities. With the SGD optimizer, this process can be expressed by
\begin{align*}
\boldsymbol{g}_{m,r}^i\leftarrow\boldsymbol{g}_{m,r}^{i-1}-\eta \nabla F_m(\boldsymbol{g}_{m,r}^{i-1},\boldsymbol{\xi}_{m,r}^{i}),\;\;\; i\in[I],
    \numberthis
    \label{eq:Local_update}
\end{align*}
where the local model $\boldsymbol{g}_{m,r}^i$ at $i$-th iteration is updated based on the gradient of the local objective function $\nabla F_m(\boldsymbol{g}_{m,r}^{i-1},\boldsymbol{\xi}_{m,r}^{i})$ evaluated on a data patc $\boldsymbol{\xi}_{m,r}^{i}$ extracted from $\mathcal{D}_m$, $\eta$ denotes the learning rate, and $r$ indexes the training round. {For clarity, the index $I$ is dropped below.} After the local training, clients transmit their local models to the PS, and PS aggregates the global model $\boldsymbol{G}(r)$ of the $r$-th round based on its received local models. If we express the aggregation at PS by a function from local models to the global model $\mathcal{S}: \{\boldsymbol{g}_{m,r}\} \rightarrow \boldsymbol{g}_{r}$, the aggregation at PS aims to the statistically recover unbiased the global model, i.e., 
\begin{align*}
\mathbb{E}\left[\mathcal{S}(\boldsymbol{g}_{1,r}, \cdots, \boldsymbol{g}_{M,r})\right]=\sum_{m=1}^M\omega_m\boldsymbol{g}_{m,r}. 
\numberthis
\label{eq:goal_at_PS}
\end{align*} 
Through iterative local training and communication, the objective in (\ref{eq:goal_FL}) is expected to be achieved.
\begin{remark}[Objective Inconsistency Induced by Heterogeneous Networks]
    The expectation $\mathbb{E}[\cdot]$ in (\ref{eq:goal_at_PS}) accounts for all stochastic sources throughout the entire training process, including the intermittent network connectivity.
    The asymmetry in channel statistics can induce a bias from the expected aggregation, i.e., the aggregation at PS will give $\sum_{m=1}^M\beta_m\boldsymbol{g}_{m,r}$ where $(\beta_1, \cdots, \beta_M)\neq (\omega_1, \cdots, \omega_M)$. Then, the function being optimized is the surrogate function $\sum_{m=1}^{M}\beta_mF_m(\boldsymbol{g}, \mathcal{D}_m)$, instead of our target in (\ref{eq:goal_FL}). 
    This will result in a mismatched convergence point from the true global model. This phenomenon is referred to as objective inconsistency, a prevalent issue for all local solvers \cite{wang2020tackling}. 
\end{remark}

\subsection{Communication Model}\label{sec:connectivity}
The links among clients can be captured by a random Bernoulli matrix $\boldsymbol{\mathcal{T}}(r)$, where each link $\tau_{mk}(r)\in \boldsymbol{\mathcal{T}}(r)$ from client $k$ to client $m$ is modeled as a binary erasure $\tau_{mk}(r)\sim \mathrm{Ber}(1-p_{mk})$ and $p_{mk}$ is the outage probability from client $k$ to client $m$. For $m=k$, $\tau_{mk}(r)=1$, as there is no transmission. The direct links from clients to PS can be captured by a Bernoulli vector $ \boldsymbol{\tau}(r)$, where each link $\tau_k$ from client $m$ to PS is modeled as a binary erasure $\tau_k\sim \mathrm{Ber}(1-p_{m})$ and $p_{m}$ is the outage probability from client $m$ to PS. These transmissions are assumed to be orthogonal, i.e., $\forall k\neq m: \tau_k\perp\tau_m$, $\forall (k_1,m_1)\neq (k_2,m_2): \tau_{k_1m_1}\perp\tau_{k_2m_2}$, $\forall k_1, m_1, k_2: \tau_{k_1m_1}\perp\tau_{k_2}$. The topics, such as interference and scheduling, are beyond the scope of this article and the downlink broadcasting is assumed to be error-free.


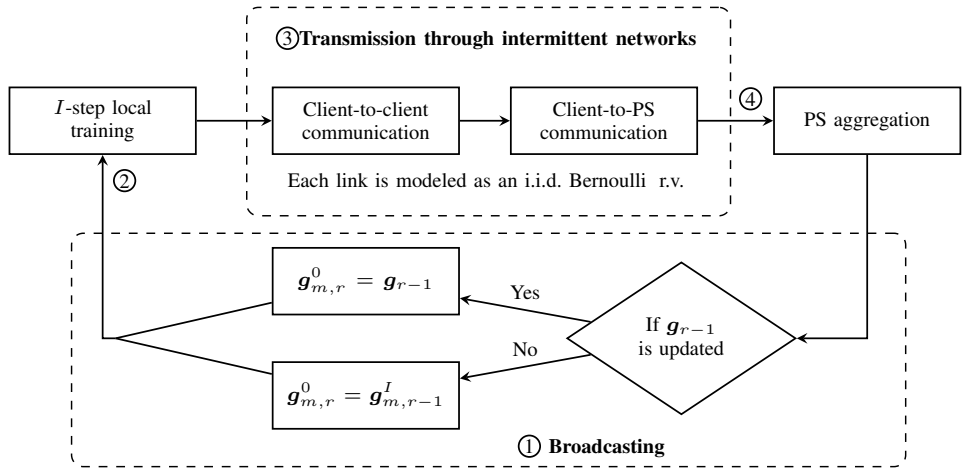
\begin{figure*}[t]
\centering
\scalebox{0.75}{\resizebox{\textwidth}{!}{
\begin{tikzpicture}[font=\scriptsize, >=stealth,nd/.style={draw,circle,inner sep=0pt,minimum size=5pt}, blkCH/.style={draw,minimum height=0.8cm,text width=2.3cm, text centered},  blk/.style={draw,minimum height=0.8cm,text width=2cm, text centered}, rounded/.style={circle,draw,minimum size=0.6cm, text centered}, diam/.style={diamond, aspect=1.5, draw, text width=4em, text centered},  x=0.6cm,y=0.55cm]
\tikzset{mult_fig/.pic={
\draw (-1,-1)--(1,1);
\draw (-1,1)--(1,-1);
}}
\tikzset{threshold_fig/.pic={
\draw (-1,0)--(1,0);
\draw[red,line width=1.0 pt] (-1,-1)--(0,-1)--(0,1)--(1,1);
}}

\path		
	(0,0,)coordinate[](q){} 
	node[blk,rounded corners,fill opacity=0, dashed, line width=0.5, minimum width=5.8cm, minimum height=2.5cm](Tx)at (9.95,-0.5){ }
        node[](Tx) at(9.9,1.1){\circled{3}\textbf{Transmission through intermittent networks}}  
        node[blk,rounded corners,fill opacity=0, dashed, line width=0.5, minimum width=10cm, minimum height=2.8cm](Tx)at (9.95,-5.7){ }
        node[](bc) at(12,-7.8){\circled{1}\textbf{ Broadcasting}} 
        node[](Tx2) at(9.9,-2){Each link is modeled as an i.i.d. Bernoulli\; r.v.}  
	node(lt)[blk, right=1 of q]{$I$-step local training}
        node(D2D)[blk, right=1.5  of lt]{Client-to-client communication} 
        node(D2P)[blk, right=1  of D2D]{Client-to-PS communication}
        node(PSag)[blk, right=1.5  of D2P]{PS aggregation}
        coordinate[right=1 of PSag](outputPS){} 
        coordinate[below=4 of D2D](judgeP){} 
        coordinate[right=5 of judgeP](judgeP1){} 
        node(success)[blk, right=1.5,above=0.5  of judgeP]{$\boldsymbol{g}_{m,r}^{0}=\boldsymbol{g}_{r-1}$}
        node(fail)[blk, right=1.5,below=0.5  of judgeP]{$\boldsymbol{g}_{m,r}^{0}=\boldsymbol{g}_{m,r-1}^I$}
        coordinate[left=5 of judgeP](judgeP2){} 
        node(judge)[diam, right=4 of judgeP]{If $\boldsymbol{g}_{r-1}$ is updated}
        
    ;

    \draw[->] (lt)--node[above, pos=0.3]{}(D2D);
    \draw[->] (D2D)--node[above]{$ $}(D2P);
    \draw[->] (D2P)--node[above, pos=0.7]{\circled{4}}(PSag);
    \draw[->] (PSag)|-node[above]{$ $}(judge);
    \draw[->] (judge)--node[above]{Yes}(success);
    \draw[->] (judge)--node[above]{No}(fail);
    \draw (success)--node[above]{$ $}(judgeP2);
    \draw (fail)--node[above]{$ $}(judgeP2);
    \draw[->] (judgeP2)-|node[right, pos=0.93]{\circled{2}}(lt);

\end{tikzpicture}
}}
\vspace{1mm}
\caption{ The training process in CoGC without recovery guarantee at the PS. } 
\label{fig:CoGC_framework}
\end{figure*}

\subsection{Gradient Coding (GC) Decoding Mechanism}
\subsubsection{{Cyclic GC Codes}}
Aiming to recover the exact sum $\boldsymbol{g}_{1,r}+\boldsymbol{g}_{2,r}+\cdots+\boldsymbol{g}_{M,r}$, GC focuses on devising a pair of structured matrices, a cyclic allocation matrix $\boldsymbol{B}$ and a combination matrix $\boldsymbol{A}$, satisfying
\begin{align*}
\boldsymbol{A}\boldsymbol{B}=\mathbf{1},
\numberthis
\end{align*}   
where $\mathbf{1}$ denotes the all-one matrix, such that 
\begin{align*}
\boldsymbol{A}\boldsymbol{B}\cdot
\begin{bNiceMatrix}
\boldsymbol{g}_{1,r}\\
\vdots\\
\boldsymbol{g}_{M,r}
\end{bNiceMatrix}
=
\begin{bNiceMatrix}
\boldsymbol{g}_{1,r}+\boldsymbol{g}_{2,r}+\cdots+\boldsymbol{g}_{M,r}\\
\vdots\\
\boldsymbol{g}_{1,r}+\boldsymbol{g}_{2,r}+\cdots+\boldsymbol{g}_{M,r}
\end{bNiceMatrix}
.
    \numberthis
\end{align*}
The $\boldsymbol{B}$ is a cyclic matrix of size $M\times M$ with $s+1$ nonzero entries per row, and $\boldsymbol{A}$ of size ${M \choose s}\times M$ encompass all possibilities with $s$ zeros per row. With such a design, GC is robust to any $s$ stragglers. The algorithm to generate such $\boldsymbol{A}$ and $\boldsymbol{B}$ pairs is given in \cite[Algorithm II]{tandon2017gradient}.

\subsubsection{{Find the Combinator}}
{Each row in $\boldsymbol{A}$ corresponds to a unique straggler pattern. The combination row for a certain straggler pattern, assuming the $f_r$-th row, should satisfy}
\begin{align*}
\boldsymbol{a}_{f_r}=\boldsymbol{a}_{f_r}\circ \boldsymbol{\tau}(r)^\top,
    \numberthis
    \label{eq:detect straggler patterns}
\end{align*}
where $\circ$ denotes Hadamard product (element-wise).

\section{The Proposed Framework: Cooperative Gradient Coding (CoGC)}  \label{sec: CoGC} 
This section introduces the training process under the CoGC framework employing the standard GC decoding mechanism, as shown in Fig. \ref{fig:CoGC_framework}. {We note that CoGC is similar to Tandon's scheme\cite{tandon2017gradient}, in the sense that both frameworks build upon cyclic gradient codes. The key difference lies in their implementation strategies. As depicted in Fig. \ref{fig:tandon_framework} and Fig. \ref{fig:CoGC_framework}, Tandon's scheme is based on dataset sharing among clients, while ours adopts a gradient-sharing paradigm.}  

\begin{itemize}[leftmargin=*, topsep=1pt, itemsep=1mm, parsep=0pt, label=$\blacktriangleright$] 
\item \textbf{Broadcasting:}
    Due to the binary characteristic of the standard GC decoding mechanism, the global model update is not guaranteed in the $r$-th training round. If the global model was successfully updated by the PS in the last training round, PS broadcasts $\boldsymbol{g}_{r-1}$ to clients; otherwise, the clients initialize with their latest local models, i.e.,
\begin{align*}
    \boldsymbol{g}_{m,r}^{0}=\begin{cases}
        \boldsymbol{g}_{r-1},\;\;\;\;\;\;\;\;\;\;\;\;\;\mathrm{if}\;\boldsymbol{g}_{r-1} \; \mathrm{is\; updated\;}\;\;\\
        \boldsymbol{g}_{m,r-1},\;\;\;\;\;\;\;\;\;\mathrm{if}\;\boldsymbol{g}_{r-1}\;\mathrm{is\; not\; updated\;},
    \end{cases}
    \numberthis
    \label{eq:local_initial}
\end{align*}
\item \textbf{Local Training:} Each client performs $I$-step SGD in (\ref{eq:Local_update}) and computes the latest local model update $\Delta\boldsymbol{g}_{m,r}=\boldsymbol{g}_{m,r}-\boldsymbol{g}_{m,r}^{0}$. 
\item \textbf{Transmission:} Each client $k$ transmits to and hears from it $s$ neighbors. 
Suppose that the transmission is done within one resource block, it is entirely captured by $\boldsymbol{\mathcal{T}}(r)$ and $\boldsymbol{\tau}(r)$, detailed below. 
\subsubsection{{Gradient-Sharing Phase}}    
Each client $k$ shares its own gradient $\Delta\boldsymbol{g}_{k,r}$ with its $s$ neighbors according to the nonzero pattern of the $k$-th column in $\boldsymbol{B}$, that is, the neighbors in the set $\mathcal{K}_1(k)=\{m\;\vert\; m\neq k: b_{mk}\neq 0 \}$. 

Each client $m$ hears from its $s$ neighbors according to the nonzero pattern of the $k$-th row in $\boldsymbol{B}$, that is, the neighbors in the set $\mathcal{K}_2(m)=\{k\;\vert\; k\neq m: b_{mk}\neq 0 \}$. Then, client $m$ computes the partial sum as 
\begin{align*}
    \boldsymbol{s}_{m,r}=\sum_{k=1}^M \hat{b}_{mk} \Delta\boldsymbol{g}_{k,r}, 
    \label{eq:partial sum}
    \numberthis
\end{align*}
where $\hat{b}_{mk}= b_{mk}\tau_{mk}(r)$ indicates whether the coding coefficient is received or not. If client $m$ successfully hears from client $k$, then $\hat{b}_{mk}= b_{mk}$. Otherwise, $b_{mk}$ is erased. 
\vspace{2mm}

If client $m$ successfully hears from all of its $s$ neighbors, i.e., $\forall k: \hat{b}_{mk}= b_{mk}$, then (\ref{eq:partial sum}) is called \textit{complete partial sum}. Otherwise, it is \textit{incomplete partial sum}. Denote the set of clients who compute the complete partial sums by $\mathcal{K}_3(r)=\{m\in [M]: \forall k\in \mathcal{K}_2(m), \hat{b}_{mk}=b_{mk}\}$and teh set of incomplete partial sums by $\mathcal{K}_3^c(r)=[M]\setminus \mathcal{K}_3(r)$.  
\vspace{2mm}
\subsubsection{{Transmitting Partial Sums}}  
By employing the standard GC decoding mechanism, it is enough to transmit only the complete partial sums in $\mathcal{K}_3(r)$ to PS. 
\item \textbf{Aggregation at PS:}
If the number of received complete partial sums exceeds $s$, PS can detect the stragglers' pattern as in (\ref{eq:detect straggler patterns}), and combines the received partial sums according to the $f_r$-th pattern and update the global model, i.e.,   
\begin{align}
    \Delta\boldsymbol{g}_r=\frac{1}{M}&\sum_{m=1}^M \boldsymbol{a}_{f_r} \boldsymbol{s}_{m,r}=\frac{1}{M}\sum_{m=1}^M \Delta\boldsymbol{g}_{m,r},\\ 
    \label{eq: global_update_at_PS}
    &\boldsymbol{g}_r\leftarrow \boldsymbol{g}_{r-1}+\Delta\boldsymbol{g}_r.
\end{align}
However, if the number of received complete partial sums is less than $s$, PS cannot compute any useful information using the standard GC decoding mechanism.
\end{itemize}
\vspace{2mm}
{
Naturally, there are two designs of the updated rule. 
\begin{design}\label{design1}
    Repeat the communication until PS successfully recovers the exact global model, then proceed to the $r+1$th round of training. In this case, the global model recovery is guaranteed each round. 
\end{design}
\begin{design}\label{design2}
    Ignore the global model update and proceed to the next round of training with the latest local models. In this design, a global model update is not guaranteed per round, and the updated round is unpredictable. Accordingly, the entire training process ends conditioned on both that the total number of training rounds reaches the preset threshold $T$ and that the global model is successfully updated in the end.
\end{design}
}
\begin{remark}[Optimality of CoGC]
In both designs, CoGC results in either perfect recovery or complete failure. Whenever PS can compute the global model, (\ref{eq:goal_at_PS}) is always met. Hence, CoGC inherently avoids the objective inconsistency issue and guarantees the optimality of learning algorithms as $T\rightarrow \infty$.
\end{remark}
\begin{remark}[{Complexity Analysis}]
{In each round, each client is required to perform two encoding operations: one for its own local model, and another for its computed partial sum, leading to a total of $2M$ encoding operations. Furthermore, each client must decode messages from its $s$ neighbors, resulting in a total of $sM$ decoding operations across the network. In terms of memory requirements, each client needs to store $s$ elements in $\boldsymbol{B}$, which are coefficients of the partial sums. }
\end{remark}

\section{Performance Analysis}
\subsection{Global Aggregation Failure}\label{sec: outage}
In this section, we generalize the outage analysis of CoGC to general heterogeneous networks \cite{weng2024supplementaryfilecooperativegradient}. 
{For clarity, let us first distinguish two types of outages in CoGC.}
\begin{itemize}
\item {Communication outage: disruption of a single link.}
\item {The overall outage: the aggregation failure at the PS, i.e., when PS receives less than $M-s$ complete partial sums. The overall outage probability $P_O$ reflects the reliability of CoGC.} 
\end{itemize}
A high number of communication outages across the network will result in the overall outage at the PS. The failure scenarios can be sorted into the following disjoint subcases. 
\begin{subcase} 
If more than $s$ clients fails to hear from all of its $s$ neighbors in the gradient-sharing phase, the overall outage will happen for sure. The probability of this event is given by  
\begin{align*}
    P_1=\hspace{-2mm}&\sum_{\substack{\mathcal{S}\in[M] \\ \lvert \mathcal{S} \rvert>s}}
    \prod_{m\in \mathcal{S}} 
    \underbrace{\hspace{-1mm}\left( 1-\hspace{-4mm}\prod_{k_1\in\mathcal{K}_1(m)} \hspace{-4mm} \left(1-p_{mk_1}\right)\right)\hspace{-1mm}}_{P_{11}} \prod_{m\notin \mathcal{S}}  \hspace{0mm}\prod_{k_1\in\mathcal{K}_1(m)} \hspace{-4mm}\left(1-p_{mk_1}\right),
    \label{eq:outage1} 
    \numberthis
\end{align*}
where $P_{11}$ is the probability that client $m$ fails to hear from all of its $s$ neighbors. 
\end{subcase}
\begin{subcase}
If all clients successfully hear from all of their neighbors in the gradient-sharing phase, the overall outage happens when at least $s$ links from clients to PS are in outage. The probability of this event is given by 
\begin{align}
    &P_2= 
    {\underbrace{%
     \vphantom{  \sum_{\substack{\mathcal{S}\in[M], \\ \lvert \mathcal{S} \rvert>s}}} 
     \prod_{m=1}^M \prod_{k_1\in\mathcal{K}_1(m)} \hspace{-3mm}\left(1-p_{mk_1}\right)}_{P_{21}}}
     \cdot     \hspace{-2mm}{\underbrace{\sum_{\substack{\mathcal{S}\in[M] \\ \lvert \mathcal{S} \rvert>s}} \prod_{ k_2\in \mathcal{S} } p_{k_2} \prod_{ k_2\notin \mathcal{S} } \left(1-p_{k_2}\right)
    }_{P_{22}}},
\end{align}
where $P_{21}$ is the probability that all communication links are in good condition during the gradient-sharing phase, and $P_{22}$ is the probability that more than $s$ links from clients to PS are in outage.
\end{subcase}
\begin{subcase}
If there are $v_1\in\{1, \cdots, s\}$ 
clients cannot hear from all of its $s$ neighbors in the gradient-sharing phase (denote this set of clients by $\mathcal{S}_1$), then the overall outage will occur when more than $s-v_1$ links from clients to PS are in outage. 
The probability that only the $v_1$ clients in $\mathcal{S}_1$ are fails to collect from all of its $s$ neighbors is given by
\begin{align*}
    P_{31}=&\prod_{m\in \mathcal{S}_1} 
     \left( 1-\hspace{-4mm}\prod_{k_1\in\mathcal{K}_1(m)} \hspace{-4mm} \left(1-p_{mk_1}\right)\right) \prod_{\substack{m\in \mathcal{S}_1^c }} \prod_{k_2\in \mathcal{K}_1(m)}\hspace{-4mm} \left(1-p_{mk_2}\right).
     \numberthis
     \label{eq:1-s,d2d} 
\end{align*}
Conditioned on the previous event, the probability that more than $s-v_1$ links (denoted the set by $\mathcal{S}_3$) from clients to the PS are in outage is given by 
\begin{align*}
P_{32}=\sum_{\substack{\mathcal{S}_3\in \mathcal{S}_1^c\\ \lvert \mathcal{S}_3 \rvert>s-v_1}}
    \hspace{-2mm}\prod_{m\in \mathcal{S}_3} \left( 1-\prod_{k_3\in\mathcal{K}_1(m)} \left(1-P_{mk_3}\right)\right)
    .
     \numberthis
\end{align*}
Thus, when the clients in $\mathcal{S}_1$ fail to hear from all of its $s$ neighbors, the overall outage probability is given by $P_{31}\cdot P_{32}$. Next, by counting all the possible patterns of $\mathcal{S}_1$ and summing up their probabilities, we get the overall outage probability for subcase 3, i.e.,
\begin{align*}
    P_3=\sum_{v_1=1}^s \sum_{\substack{\lvert \mathcal{S}_1\rvert=v_1\\ \mathcal{S}_1\subset [M]}} P_{31}\cdot P_{32}.
    \numberthis
\end{align*}
\end{subcase}
Since these subcases are non-overlapping, the overall outage probability in CoGC is given by
\begin{align*}
    P_O=P_1+P_2+P_3.
    \numberthis
    \label{eq:outageall}
\end{align*}
From previous analysis, we learn that $P_O$ is fully determined by $s$ and the statistics of $\boldsymbol{\mathcal{T}}(r)$ and $\boldsymbol{\tau}(r)$. In fact, $P_O$ not only serves as a bridge between individual communication outages and system performance, but also provides information on CoGC utilization.
\begin{remark}[Repeated Rounds Before Success]\label{remark:commu. attempts}
Given that the overall outage probability $P_O$ represents the likelihood of global model recovery failure in a single attempt, and assuming independence of communications across training rounds, the number of communication rounds between two consecutive successful recoveries, denoted by $R_r$, follows a geometric distribution, i.e., $R_r\sim\mathrm{Geo}(1-P_O)$. Accordingly, the expected number of rounds between two successful recoveries is 
\begin{align}
    \mathbb{E}[R_r]=\frac{1}{1-P_O}.
    \label{eq: number_com}
\end{align}
The (\ref{eq: number_com}) shows that a large outage probability $P_O$ results in a high expected number of communication attempts before a successful global model recovery. In Design \ref{design1}, with a recovery guarantee of the global model recovery, large $P_O$ indicates a tremendous waste of communication resources. {In Design \ref{design2}, without a recovery guarantee, the number of consecutive local training iterations $R_rI$ is expected to be large. This may lead to the accumulated divergence among local models obtained from consecutive local training. This divergence, in turn, increases the variance of the aggregated global model at the PS, implied by \cite[Lemma 2 and 3]{weng2024cooperative}, thereby hindering convergence. As a consequence, achieving optimal distributed training performance requires a considerably large total number of training rounds $T$.} 
\end{remark}

\begin{remark}[A Case Study on Large $P_O$ Conditions]\label{remark:large Po}
If $p_{mk}$ is relatively large for $\forall m$, $\forall k$, the probability that client $m$ cannot hear from all of its neighbors, $P_{11}$, will become considerable. As a result, the outage probability $\prod_{m\in[M]}P_{11}$ corresponding to the event that all $M$ clients fail to hear from all of their neighbors is large, it can even approach $1$. For instance, when $p_{mk}=0.4$, $M=10$, $s=7$, then $\prod_{m\in[M]}P_{11}=0.7528$, leading to a substantially high overall outage probability $P_O$. This example suggests that CoGC is unsuitable for scenarios where communications among clients are in poor condition in general. {This conclusion is further supported by Fig. \ref{fig:Overall_outage}.}
\begin{figure}[!htp]
\hfill
\vspace{-2mm}
   \begin{minipage}[b]{0.49\textwidth}
  \centering
   \includegraphics[width=0.7\linewidth]{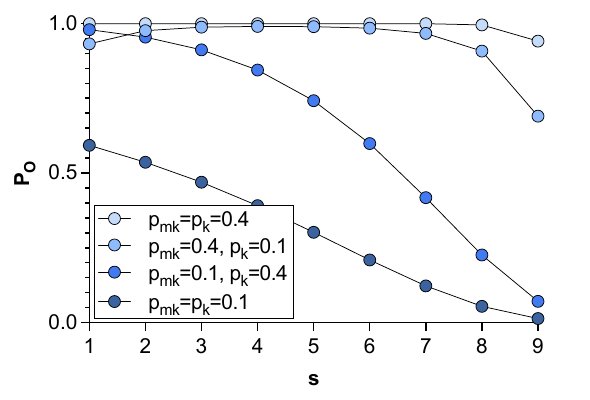}    
    \caption{{The failure probability of global model reconstruction $P_O$ in terms of $s$ in several study cases.} }
    \label{fig:Overall_outage}
    \end{minipage}
    \vspace{-8mm}
\end{figure}
\end{remark}

\subsection{Convergence Analysis} 
For Design \ref{design1}, CoGC with a recovery guarantee, the convergence is guaranteed \cite{raviv2020gradient}. This section extends our previously non-convex convergence analysis in \cite{weng2024supplementaryfilecooperativegradient} for Design \ref{design2} by relaxing the condition on $T$ from infinite to large yet finite. The non-convex convergence analysis can generally provide insights into modern deep neural networks (DNNs).

To begin with, we state several assumptions on the local objective function to ensure a feasible analysis.
\begin{assumption}[{$L$-Smoothness}]
    Each local objective function is bounded by $F_m(x)\geq F^\star$ and is differentiable, its gradient $\nabla F_m(x)$ is L-smooth, i.e., $\lVert \nabla F_m(x)-\nabla F_m(y) \rVert\leq L\lVert x-y \rVert$, $\forall i \in [M]$.
\end{assumption}
\begin{assumption}[Unbiasedness and Bounded Variance]
   The local stochastic gradient is an unbiased estimation, i.e., $\mathbb{E}_\xi[\nabla F_m(x,\xi)]=\nabla F_m(x)$, and has bounded data variance $\mathbb{E}_\xi[\lVert \nabla F_m(x,\xi)-\nabla F_m(x)\rVert^2]\leq \sigma^2 $, $\forall i \in [M]$. 
\end{assumption}
\begin{assumption}[Bounded Heterogeneity]
    The dissimilarity between $\nabla F_m(x)$ and $\nabla F(x)$ is bounded, i.e., $\lVert \nabla F_m(x)-\nabla F(x)\rVert^2\leq D_m^2$, $\forall i \in [M]$. 
\end{assumption}
Based on assumptions 1--3, and Lemmas 2--4 in \cite{weng2024supplementaryfilecooperativegradient}, we derive a convergence bound with a probability guarantee of $99.86\%$, formally stated below.
\begin{theorem}\label{theo:convergence CoGC}
Given $I$ local training iterations, for any algorithm that results in binary recovery of the global model with failure probability of $P_O$ per round and updates the global model as in (\ref{eq:local_initial}), if the learning rate is set as $\eta=\frac{1}{L}\sqrt{\frac{M}{T}}$ and the number of training rounds $T$ is  is sufficiently large but finite, the following convergence bound holds with probability more than $99.86\%$, 
\begin{align*}
    &\min_{r\in [T]} \mathbb{E}\left[\left\lVert\nabla F(\boldsymbol{g}_{r}^{0}) \right\rVert^2\right] 
    \\
    &\leq\frac{\mu_{\mathcal{J}_2}}{\mu_{\mathcal{J}_1}}+3\left( \frac{\sigma_{\mathcal{J}_2}^2}{\mu_{\mathcal{J}_1}^2T}+\frac{\mu_{\mathcal{J}_2}^2\sigma_{\mathcal{J}_1}^2}{\mu_{\mathcal{J}_1}^4T}+2\frac{\mu_{\mathcal{J}_2}\sigma_{\mathcal{J}_1} \sigma_{\mathcal{J}_2}}{\mu_{\mathcal{J}_1}^3 T} \right)\overset{\triangle}{=}\epsilon(P_O),
    \numberthis
    \label{eq_theorem1}
\end{align*}
where $\mu_{\mathcal{J}_1}$, $\mu_{\mathcal{J}_2}$, $\sigma_{\mathcal{J}_1}$ and $\sigma_{\mathcal{J}_2}$ are given in (\ref{eq:E_J1}), (\ref{eq:E_J2}), (\ref{eq:sigma1^2}), and (\ref{eq:sigma2^2}) respectively. 
\end{theorem}
\begin{proof}
    Proof of Theorem 1 is provided in Appendix \ref{appx: Proof of Theorem 1}. 
\end{proof}
\begin{remark}
    Given the choice of $\eta=\frac{1}{L}\sqrt{\frac{M}{T}}$, the optimility gap converges to $0$ with rate $\mathcal{O}\left(\frac{1}{\sqrt{T}}\right)$.
\end{remark}
\begin{remark}\label{remark: convergence wrt data hetero}
    From \cite[Theorem 1]{weng2024cooperative}, the convergence bound is magnified by $\frac{1}{1-P_O}\sum_{k=1}^K\omega_m D_m^2$ if $P_O\rightarrow 1$.
\end{remark}
\subsection{Secure Aggregation at PS}
The GC decoding mechanism only allows PS to compute the global model by aggregating the partial sums, instead of learning about each individual, thus achieving secure aggregation.
This section examines the secure aggregation feature of the GC decoding mechanisms by context-dependent local mutual information privacy (CD-LMIP) \cite{9448019}, reflecting how much PS can learn about each local model from each complete partial sum. In the analysis below, $\boldsymbol{B}$ is known to PS, $\boldsymbol{g}_m\in \mathbb{R}^d$ is assumed to be a Gaussian vector, as the true local model is corrupted by several noise sources, whose superposition is approximated by Gaussian noise due to the Central Limit Theorem (CLT) \cite{saha2022minimaxoptimalquantizationlinear}. Since the local training is performed independently on each client, $\boldsymbol{g}_{m}$ is also assumed to be independently distributed. 
\begin{definition}[CD-LMIP]
    For a distribution space $P_{\mathbf{X}}$, a randomized mechanism $\mathcal{M}$ satisfies $\mu$-CD-LMIP iff
    \begin{align}
        \mathrm{I}(\mathbf{X}; \mathcal{M}(\mathbf{X}))\leq \mu\;\;\mathrm{bits}
    \end{align}
\end{definition}

\begin{lemma}[LMIP in Complete Partial Sums]\label{theorem3}
    For mutually independent Gaussian vectors $\boldsymbol{g}_{m}$ with covariance matrix $\Sigma^2_{m}$, $\forall m\in [M]$, the complete partial sum $\sum_{k=1}^M b_{mk}\boldsymbol{g}_{k}$ satisfy $\mu$-CD-LMIP, where 
    \begin{align}
        \mu=\frac{d}{2}\log\frac{\det\left(\sum_{k=1}^M b_k^2 \Sigma_k\right)}{\det\left(\sum_{k\neq m} b_k^2 \Sigma_k\right)} \;\;\mathrm{bits}.
    \end{align}
\end{lemma}
\begin{proof}[Proof Sketch ]
Lemma \ref{theorem3} can be easily proved by the basic, well-known properties of mutual information (MI).
\end{proof}

\section{Cost-efficient Cyclic GC Codes}\label{sec: Cost-eff GC} 
The $s+1$ nonzeros in each row of $\boldsymbol{B}$ decide the cost, either the computational load in the dataset-replication-based GC framework, or the communication overhead in the gradient-sharing-based GC framework. Following the last section, where we quantified the failure probability of the standard GC decoding mechanism, this section formulates an optimization problem to determine the most cost-efficient choice of $s$ while ensuring the $P_O$ meets the requirement.
\subsubsection{Communication Costs in CoGC}\label{sec:num_comm_cost}
In the gradient-sharing phase, each client communicates with its $s$ neighbors, leading to a total of $sM$ transmissions. depending on whether each client fails to receive messages from all of its neighbors responsible for communicating with the PS. Thus, there are at most $(s+1)M$ transmissions in total, corresponding to the scenario where no communication outage occurs in the gradient-sharing phase.    
\subsubsection{Cost-efficient GC Design} Given a target level of $P_O^*$ and the network statistics, $p_m$ and $p_{mk}$, the most cost-efficient GC design with $s^*$ nonzero entries per row is achieved by selecting the smallest $s$ that results in an outage probability lower than $P_O^*$, i.e.,  
\begin{subequations}
    \begin{align}
    s^*=&\underset{s}{\mathrm{arg\,min}} \, P_O(s),\\
    &\mathrm{s.t.} \quad P_O(s) \leq P_O^*. 
    \label{eq: min_c^jluster}
\end{align}
\label{eq: cost-eff GC}
\end{subequations}
Notably, $P_2$ decreases monotonically with $s$, whereas $P_1$ and $P_3$
do not necessarily follow this trend. Their behavior depends on the joint effect of the number of consecutive multiplications of communication opportunities (both successes and outages) and the number of possible sets, both of which are not necessarily monotonous with $s$. As the combined effect, $P_O$, being the sum of $P_1$, $P_2$, and $P_3$, is not necessarily a monotonous function with $s$. Its behavior depends on specific values of $p_m$ and $p_{mk}$. Fortunately, given $p_m$ and $p_{mk}$, $P_O(s)$ can be computed under different choice of $s$ according to the close-form formula given in (\ref{eq:outage1})-(\ref{eq:outageall}), making (\ref{eq: cost-eff GC}) always solvable. 
\section{GC$^+$: Enhanced Reliability by\\ Recycling the Wastes} \label{sec: GC$^+$}  If more than $s$ incomplete partial sums are present at the PS, it is impossible to compute the exact sum by the standard GC decoding mechanism. However, instead of discarding these incomplete partial sums, leveraging their remaining information can be valuable for reconstructing the global model. This section introduces such a decoding mechanism, named GC$^+$, which enhances the original GC decoding approach by integrating a complementary decoding strategy.
{The GC$^+$ improves straggler tolerance by leveraging the inherent structure of perturbed coefficient matrices caused by communication disruptions. This improvement can even enable retrievals of all local models and recover the exact global model, possibly within one complete transmission round in CoGC.
}

\vspace{-0.5em}
\subsection{The GC$^+$ Decoding}
Below, we present the GC$^+$ decoding mechanism under the CoGC framework, highlighting its differences from the standard GC decoding mechanism. The complete training procedure is outlined in Algorithm \ref{alg:GC$^+$}, while the GC$^+$ is detailed in Algorithm \ref{algo:GC$^+$_decoding}\footnote{Algorithm \ref{algo:GC$^+$_decoding} is an approximate detection of the full-rank blocks with significantly reduced complexity. Due to dominance by full recovery, the approximation is sufficiently good. } (in MATLAB syntax).
\begin{itemize}[leftmargin=*, topsep=1pt, itemsep=1mm, parsep=0pt, label=$\blacktriangleright$] 
    \item \textbf{Broadcasting}: At the beginning of the $r$-th round training, PS broadcasts the latest global model $\boldsymbol{g}_{r-1}$ to all clients.     
    \item The \textbf{Local Training} follows standard procedures in (\ref{eq:Local_update}).
    \item \textbf{Communication:} For ease of reliability analysis, a fixed number of repeated communications, $t_r$, is assumed. However, it is worth noting that the communication can also be repeated until PS can decode a subset of individuals without a preset threshold. The gradient-sharing phase is identical to Section \ref{sec: CoGC}. 
However, to enable the GC$^+$ mechanism, both the complete and incomplete partial sums need to be transmitted to the PS. 
In each communication round, say $i_r$-th, $i_r\in [t_r]$, the GC matrices pair, $\boldsymbol{A}_{i_r}$ and $\boldsymbol{B}_{i_r}$, is generated independently. Through the intermittent networks, the coefficients of the partial sums, $\hat{\boldsymbol{B}}_{i_r}$, received at PS, is given by 
\begin{align*}
\hat{\boldsymbol{B}}_{i_r}=\left(\boldsymbol{B}_{i_r} \circ \boldsymbol{\mathcal{T}}_{i_r}(r)\right)\bullet\boldsymbol{\tau}_{i_r}(r),
\numberthis
\label{eq: perturb_B}
\end{align*}
where $\bullet$ denotes the row-wise face-splitting product, and $\boldsymbol{\tau}_{i_r}(r)$ and $\boldsymbol{\mathcal{T}}_{i_r}(r)$ follows a similar definition as $\boldsymbol{B}_{i_r}$. 
\item \textbf{The GC$^+$ Decoding:}  
If PS manages to collect at least $n-s$ complete partial sums in \textbf{any} communication round, it computes the global model using the standard GC decoding mechanism in (\ref{eq: global_update_at_PS}).
Otherwise, the PS applies the following complementary decoding mechanism. First, PS identifies the largest determined or overdetermined submatrix, $\hat{\boldsymbol{B}}_{\mathrm{sub}}$, within the vertically stacked received coefficient matrices in the $r$-th training round, $\hat{\boldsymbol{B}}(r)=[\hat{\boldsymbol{B}}_{1}; \cdots; \hat{\boldsymbol{B}}_{t_r}]$. 
The $\hat{\boldsymbol{B}}_{\mathrm{sub}}$ can be formed by excluding certain rows in $\hat{\boldsymbol{B}}(r)$. The detection of $\hat{\boldsymbol{B}}_{\mathrm{sub}}$ is outlined in Algorithm \ref{algo:detect_hatB}. If such $\hat{\boldsymbol{B}}_{\mathrm{sub}}$ exists, the corresponding local models, denoted by the set $\mathcal{K}_4(r)$, can be recovered by solving the matrix equation associated with the received partial sums. The global model is then updated as 
\begin{align*}
   \boldsymbol{g}_r=\frac{1}{\lvert \mathcal{K}_4(r)\rvert} \sum_{m\in \mathcal{K}_4(r)} \boldsymbol{g}_{m,r}.
   \numberthis
   \label{eq: GullGC_PSagg}
\end{align*}
As will be seen in Fig. \ref{fig: count for full recovery}, the probability of the PS decoding nothing is low. Thus, it is reasonable to condition the global model update on the non-empty set of decoded local models when GC$^+$ applies. The communication process repeats until the PS can decode at least one local model. 
\end{itemize}

\begin{algorithm}[t]
\caption{Training Process Using GC$^+$ in CoGC}\label{alg:GC$^+$}
\KwIn{$T$}
\KwOut{$\boldsymbol{g}_r$}
\textbf{Initialize:} $\boldsymbol{ g}_0 \gets 0$; $r \gets 1$; $\mathcal{K}_4(r), \hat{\boldsymbol{B}}(r)\gets\emptyset$; $t_r\gets 0$;\\
\While{$r\leq T$}{
    PS broadcasts $\boldsymbol{g}_{r-1}$ to all clients\;
    \For{$m=1,\cdots,M$}{
    Client $m$ sets $\boldsymbol{g}_{m,r}^{0}=\boldsymbol{g}_{r-1}$\;
    Client $m$ performs $I$-step SGD as (\ref{eq:Local_update})\;
    }
    \While{$\mathcal{K}_4(r)=\emptyset$}{
    \For{$i_r=1, \cdots, t_r$}{
    \For{$m=1,\cdots,M$}{
    Client $m$ sends  $b_{km}^{i_r} \boldsymbol{g}_{m,r}$ to client $k\in\mathcal{K}_1(m)$\;
    Client $m$ hears from clients in $\mathcal{K}_2(m)$\;
    Client $m$ computes and transmits (\ref{eq:partial sum}) to PS\;
    $\hat{\boldsymbol{B}}(r)=[\hat{\boldsymbol{B}}(r); \hat{\boldsymbol{B}}_{i_r}]$
    }}
    PS performs GC$^+$ decoding as in Algorithm \ref{algo:GC$^+$_decoding}\;
    }
    $r=r+1$;
    }  
\end{algorithm}
\begin{remark}
    While GC$^+$ provides flexibility and enhanced resilience against stragglers, it compromises privacy at the PS, as the PS can decode individual local models. However, it remains secure against potential eavesdroppers during client-to-PS communication. 
    {To further address privacy concerns, GC$^+$ can be seamlessly combined with e.g., the Gaussian mechanism, enhancing privacy preservation. }
\end{remark}

\begin{algorithm}[t]
\caption{Detection of $\hat{\boldsymbol{B}}_{\mathrm{sub}}$}\label{algo:GC$^+$_decoding}\label{algo:detect_hatB}
\KwIn{$\hat{\boldsymbol{B}}(r)$ of size $Mt_r \times M$}
\KwOut{$\hat{\boldsymbol{B}}_{\mathrm{sub}}(r)$, $\mathcal{K}_4(r)$, $\boldsymbol{g}_r$}
\eIf{$\forall i_r\in [t_r]: \hat{\boldsymbol{B}}_{i_r}$ contains $s$ complete rows}{
PS computes the global model as in (\ref{eq: global_update_at_PS}) \;
$\mathcal{K}_4(r)=[M]$\;
}{
$\boldsymbol{E}(r)=\mathrm{rref}(\hat{\boldsymbol{B}}(r))$\;
\Comment{Compute the reduced row Echelon form of $\hat{\boldsymbol{B}}(r)$}
$\mathcal{K}_4(r)\defeq\{i\in [M]: \boldsymbol{e}(:,i)\in\boldsymbol{E}(r), \boldsymbol{e}(:,i)\neq \boldsymbol{0}_{Mt_r\times 1}\}$\;
$\mathcal{K}_5(r)\defeq\{i\in [Mt_r]:\boldsymbol{e}(i,:)\in\boldsymbol{E}(r), \boldsymbol{e}(i,:)\neq \boldsymbol{0}_{1\times M}\}$\;
\Comment{Detect the nonzero columns and rows in $\boldsymbol{E}(r)$}
\eIf{$\lvert \mathcal{K}_4(r) \rvert<\lvert \mathcal{K}_5(r) \rvert $}{
$\{\boldsymbol{g}_{k,r}: k\in \mathcal{K}_4(r)\} $ can be solved using the associated received partial sums\;
PS computes the global model as in (\ref{eq: GullGC_PSagg});\\
}{
$\mathcal{K}_4(r)=\emptyset$\;
}
}
\end{algorithm}

\vspace{-0.5em}
\subsection{Reliability Analysis}\label{GC$^+$: Outage analysis}
{Unlike the standard GC decoding mechanism, which solely relies on the all-one vector in the subspace of the gradient codes, GC$^+$ additionally leverages the rank of the received coding coefficient matrices. The GC$^+$ demonstrates significantly improved reliability for two main reasons. First, the vertical concentration increases the rank of the received coefficient matrix. Second, and somewhat anti-intuitively, the client-to-client links outages increase the rank of each $\hat{\boldsymbol{B}}_{i_r}$.} This section presents a rigorous analysis of the rank of the coding coefficients in each learning phase and evaluates the reliability of GC$^+$ by deriving bounds on the probability of successfully recovering all local models.
\subsubsection{{Rank Increase by Outages}}
We observe that although $\mathrm{rank}(\boldsymbol{B}_{i_r})$ is less than $M$, its perturbed counterpart $\hat{\boldsymbol{B}}_{i_r}$, given in (\ref{eq: perturb_B}), is likely to achieve higher rank, even full rank. {This is attributed to random network outages among clients, which destroy the structured design of the cyclic GC codes but induce beneficial rank increases that improve the likelihood of successful recovery in GC$^+$.} In fact, the client-to-client outages consistently contribute to an increased rank of the coefficient matrix. The strict rank enhancement is stated in the following lemma. 
\begin{lemma}[Rank Enhancement by Client-to-client Outages]\label{Lemma_rank_increase_d2d}
Define the matrix of perturbed coefficients by client-to-client outages as $\tilde{\boldsymbol{B}}_{i_r}=\boldsymbol{B}_{i_r}\circ \boldsymbol{\mathcal{T}}_{i_r}(r)$. 
We have $\mathrm{rank}(\tilde{\boldsymbol{B}}_{i_r})\geq M-s$ w.p. $1$.
More specifically, if the number of the unperturbed rows in $\boldsymbol{\mathcal{T}}_{i_r}(r)$ exceeds $M-s$, the rank of $\tilde{\boldsymbol{B}}_{i_r}$ is given by  
\begin{align*}
\mathrm{rank}(\tilde{\boldsymbol{B}}_{i_r})=\min\{ M, M-s+n_{i_r}\},  
\numberthis
\label{eq: rank_D2D}
\end{align*}
where $n_{i_r}$ is given in Appendix \ref{appendix: rank_analysis}. 
\end{lemma}
\begin{proof}[Proof Sketch]
 First, we show that any coefficient matrix $\boldsymbol{B}_{i_r}$ in GC codes has rank $M-s$ w.p. $1$. Consider $\boldsymbol{H}_{i_r}(r)\in \mathbb{R}^{s\times M}$ constructed in \cite[Algorithm II]{tandon2017gradient}, and the subspace $\mathcal{S}_{i_r}$ defined by its null space is given by 
\begin{align*}
    \mathcal{S}_{i_r}=\{\boldsymbol{x}\in \mathbb{R}^{M}: \boldsymbol{H}_{i_r}(r)\boldsymbol{x}=\boldsymbol{0}\}.
    \numberthis
\end{align*}
It can be inferred that $\boldsymbol{H}_{i_r}(r)$ is of rank $s$, since any $s$ columns in $\boldsymbol{H}_{i_r}(r)$ are linearly independent w.p. $1$ and that $\mathrm{rank}(\boldsymbol{H}_{i_r}(r))\leq \min\{s, M\}=s$. Thus, the dimension of its null space is $\dim(\mathcal{S}_{i_r})=M-s$. Moreover, every row in $\boldsymbol{B}_{i_r}$ lies in the subspace $\mathcal{S}_{i_r}$, implying $\mathrm{rank}(\boldsymbol{B}_{i_r})\leq M-s$, as the independent dimensions spanned by these rows are limited by $\mathcal{S}_{i_r}$. Furthermore, any $M-s$ rows in $\boldsymbol{B}_{i_r}$ are linearly independent w.p. $1$, which implies $\mathrm{rank}(\boldsymbol{B}_{i_r})\geq M-s$. Thus, $\mathrm{rank}(\boldsymbol{B}_{i_r})= M-s$. The supplementary supporting proofs can be found in \cite{tandon2017gradientcoding}. 

Now we proceed to analyze the rank of the perturbed coefficient matrix $\tilde{\boldsymbol{B}}_{i_r}$. If we extract an $(M-s)\times(M-s)$ sub-block from any $M-s$ consecutive rows of $\tilde{\boldsymbol{B}}_{i_r}$, aligned along its diagonal, 
the inherited structure of $\tilde{\boldsymbol{B}}_{i_r}$ ensures that this sub-block is upper triangular with a nonzero determinant, thereby guaranteeing full rank. This holds because the $0$s and diagonal elements remain unperturbed, and the diagonal elements of $\tilde{\boldsymbol{B}}_{i_r}$ are nonzero w.p. $1$. Thus, any $M-s$ consecutive perturbed rows are of rank $M-s$, further implying that $\mathrm{rank}(\tilde{\boldsymbol{B}}_{i_r})\geq M-s$. The rank formula in (\ref{eq: rank_D2D}) is further analyzed in Appendix \ref{appendix: rank_analysis}.
\end{proof}
\subsubsection{{Rank Increase by Vertical Concentration}}
Let $\boldsymbol{B}(r)$ be the vertically stacked unperturbed coefficient matrices of the partial sums transmitted within $t_r$ rounds, i.e., $\boldsymbol{B}(r)=[\boldsymbol{B}_{1}; \cdots; \boldsymbol{B}_{t_r}]$. {The enhanced straggler tolerance of $\boldsymbol{B}(r)$ comes from two aspects. First, each $\boldsymbol{B}_{i_r}$ can independently handle stragglers. Second, the rank of $\boldsymbol{B}(r)$ is enhanced by the vertical concentration and the random nature of $\boldsymbol{B}_{i_r}$.} 
\begin{lemma}[Rank Enhancement by Vertical Concentration]\label{lemma:vertical concentartion}
    Let $\boldsymbol{B}(r)$ be the vertically concentrated form of $t_r$ unperturbed coefficient matrices $\boldsymbol{B}_{i_r}$ in GC$^+$, where $i_r\in\{1, \cdots, t_r\}$. Then, the rank of $\boldsymbol{B}(r)$ is given by 
\begin{align*}
    \mathrm{rank}\left(\boldsymbol{B}(r)\right)=\min\left\{(M-s-1)t_r+1, M\right\}.
    \numberthis
    \label{eq: rank_hatB}
\end{align*}
The vertical concentration will never reduce the matrix rank. 
\end{lemma}
\begin{proof}[Proof Sketch]
   Consider the vertical concentration $\boldsymbol{B}(r)$ of the matrices $\hat{\boldsymbol{B}}_{i_r}$, $i_r=1, \cdots, t_r$. With this construction, $\boldsymbol{B}(r)$ lies in the sum of subspaces $\sum_{i_{r}=1}^{t_r}\mathcal{S}_{i_r}=\left\{\sum_{i_{r}=1}^{t_r}\boldsymbol{x}_{i_r}: \boldsymbol{x}_{i_r}\in \mathcal{S}_{i_r}, i_r=1, \cdots, t_r \right\}$. Notably, $\boldsymbol{1}_{M\times 1}\in \mathcal{S}_{i_r}, \forall i_r\in [t_r]$, since $\boldsymbol{H}_{i_r} (:, M)=-\sum_{j=1}^{M-1} \boldsymbol{H}_{i_r} (:, j)$ in the GC code construction. Combining this with the fact that $\mathrm{rank}(\hat{\boldsymbol{B}}_{i_r})=M-s$ and that each $\hat{\boldsymbol{B}}_{i_r}$ is randomly generated from real-valued distribution over $\mathbb{R}$, we have $\dim (\sum_{i_{r}=1}^{t_r}\mathcal{S}_{i_r})=(M-s-1)t_r+1$ w.p. $1$ if $(M-s-1)t_r+1<M$, and that $\dim (\sum_{i_{r}=1}^{t_r}\mathcal{S}_{i_r})=M$ w.p. $1$ if $(M-s-1)t_r+1\geq M$. 
\end{proof}
\subsubsection{{Success Bound on Full Recovery}} \label{sec: Success Bound}
In the previous sections, we quantified the effects of multiple transmissions and client-to-client outages. This section integrates all these factors, including client-to-PS outages, and analyzes the success bound that PS can decode each local model. For simplicity, it is assumed that $p_{mk} = p_m = p$ in the analysis below.

For later use, we define auxiliary matrices $\grave{\boldsymbol{B}}(r)$ and $\check{\boldsymbol{B}}(r)$. The $\grave{\boldsymbol{B}}(r)=[\boldsymbol{B}_{1}\bullet\boldsymbol{\tau}_{1}(r); \cdots; \boldsymbol{B}_{t_r}\bullet\boldsymbol{\tau}_{t_r}(r)]$. The matrix $\grave{\boldsymbol{B}}(r)$ is obtained from $\boldsymbol{B}(r)$ by erasing the rows according to $\boldsymbol{\tau}_{i_r}(r), i_r \in \{1, \ldots, t_r\}$. Note that $\boldsymbol{\tau}_{i_r}(r)$ remains the same as that shapes $\hat{\boldsymbol{B}}(r)$ in (\ref{eq: perturb_B}). 
Additionally, $\check{\boldsymbol{B}}(r)$ is constructed by extracting $M-s$ columns from each $\boldsymbol{B}_{i_r}\bullet \boldsymbol{\tau}_{i_r}(r)$ and concatenating them vertically into a single matrix.
We also define several probabilities related to the decoding events. For clarity, these notations are summarized in Table~\ref{tab:probabilities} below. 
\begin{table}[htbp]
\centering
\caption{{Summary of decoding-related probabilities.}}
\label{tab:probabilities}
\begin{tabular}{cl}
\toprule
\textbf{Symbol} & \textbf{Description} \\
\midrule
$\grave{P}_m$ & \makecell[l]{Probability of decoding $m$ local models by applying only\\ the complementary decoding mechanism to $\grave{\boldsymbol{B}}(r)$.} \\
\midrule
$\hat{P}_m$ & \makecell[l]{Probability of decoding $m$ local models by applying only\\ the complementary decoding mechanism to $\hat{\boldsymbol{B}}(r)$.} \\
\midrule
$\check{P}_m$ & \makecell[l]{Probability of decoding $m$ local models by applying only\\ the complementary decoding mechanism to $\check{\boldsymbol{B}}(r)$.} \\
\midrule
$\hat{P}_{\mathrm{partial}}$ & \makecell[l]{Probability of decoding $1\sim M-1$ local models by\\ applying GC$^+$ to $\hat{\boldsymbol{B}}(r)$.} \\
\midrule
$\hat{P}_{\mathrm{full}}$ & \makecell[l]{ Probability of updating the global model based on all local\\ models by applying GC$^+$ to $\hat{\boldsymbol{B}}(r)$.} \\
\midrule
$\hat{P}$ & \makecell[l]{ Probability of decoding at least one local models by\\applying GC$^+$ to $\hat{\boldsymbol{B}}(r)$. Otherwise, GC$^+$ fails.}  \\
\bottomrule
\end{tabular}
\end{table}

{Based on the definitions of these probabilities, we have the following relationships,}
{\begin{subequations}
    \begin{align}
       &\hat{P}_{\mathrm{partial}}=\sum_{m=1}^{M-1} \hat{P}_m,\\
       &\hat{P}_{\mathrm{full}}\geq \max\{P_M, 1-P_O\},\\
       &\hat{P}=\hat{P}_{\mathrm{partial}}+\hat{P}_{\mathrm{full}}.
    \end{align}
    \label{eq: prob_relation}
\end{subequations}}
{Based on the construction of the auxiliary matrices and relations in (\ref{eq: prob_relation}), we now analyze the ranks and derive several useful relations on the associated decoding probabilities. }
\begin{figure}[!htb]
 \centering
\includegraphics[width=1\linewidth]{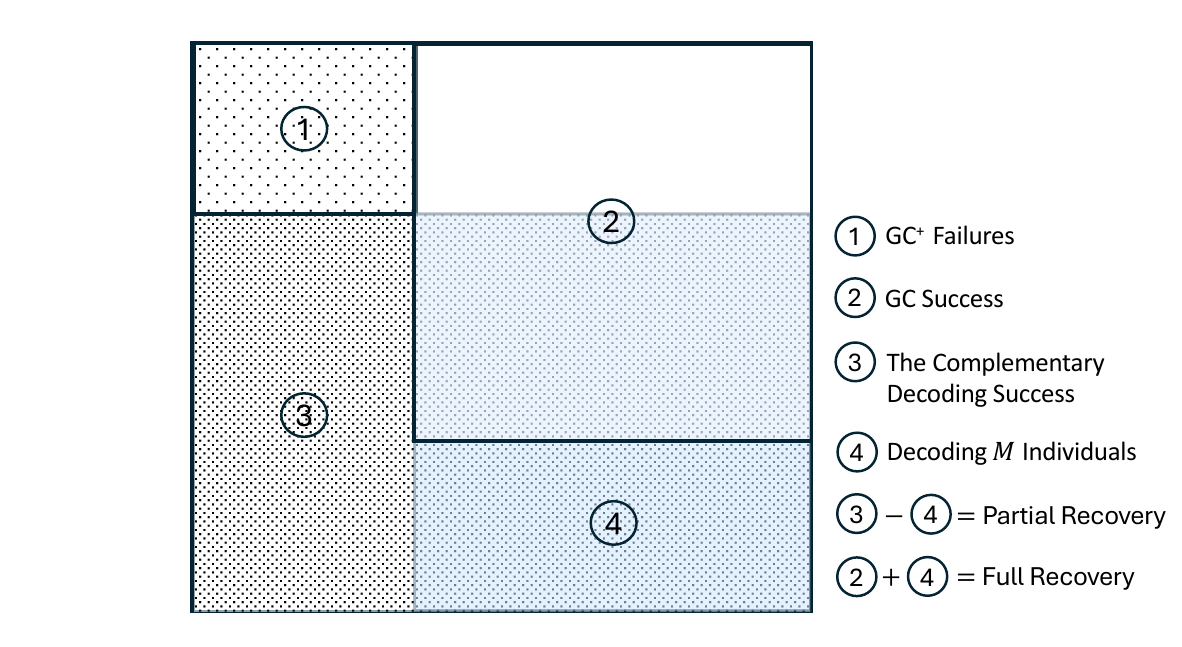}
    \caption{Illustration of partial and full recovery of the global model in GC$^+$.}
    \label{fig:outage_relation}
\end{figure}

\begin{figure}[htbp]
    \centering
    \includegraphics[width=0.75\linewidth]{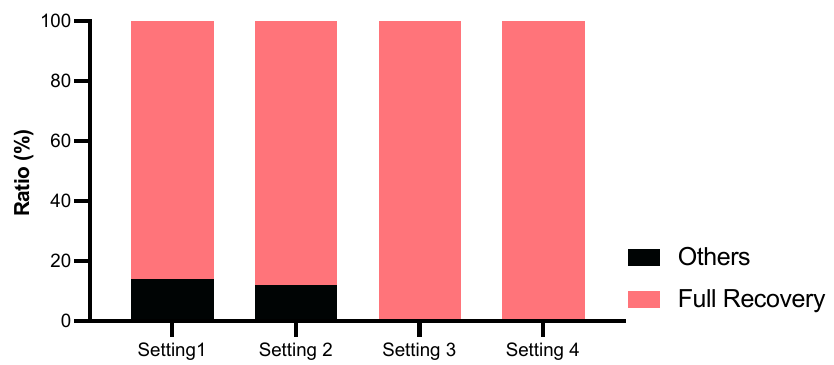}
    \caption{{Statistics of the full recovery of the global model in GC$^+$ across different networks, with $t_r=2$, $M=10$, and $s=7$. In setting 1, $p_m=0.4$, $p_{mk}=0.25$; In setting 2, $p_m=0.4$, $p_{mk}=0.5$; In setting 3, $p_m=0.75$, $p_{mk}=0.5$; In setting 4, $p_m=0.75$, $p_{mk}=0.8$.}}
    \label{fig: count for full recovery}
    \vspace{-5mm}
\end{figure}

{Following Lemma \ref{Lemma_rank_increase_d2d}, we have
\begin{align*}
    \forall \boldsymbol{\mathcal{T}}_{i_r}(r),\boldsymbol{\tau}_{i_r}(r):  \mathrm{rank}(\tilde{\boldsymbol{B}}_{i_r})\geq\mathrm{rank}(\boldsymbol{B}_{i_r}),
\end{align*} 
as client-to-client outages only contribute to a higher rank while maintaining the same $\boldsymbol{\tau}_{i_r}(r)$. Consequently, their vertical concentration satisfies 
\begin{align*}
    \forall \boldsymbol{\mathcal{T}}_{i_r}(r),\boldsymbol{\tau}_{i_r}(r):  \mathrm{rank}(\hat{\boldsymbol{B}}(r))\geq\mathrm{rank}(\grave{\boldsymbol{B}}(r)).
\end{align*} 
A direct benefit of the higher rank is that $\hat{P}_M \geq \grave{P}_M$. This follows from the fact that, for any realization of $\boldsymbol{\tau}_{i_r}(r), i_r \in \{1, \ldots, t_r\}$,  if $\grave{\boldsymbol{B}}(r)$ allows for decoding all $M$ individuals, then the higher-rank matrix $\hat{\boldsymbol{B}}(r)$ must also support the recovery of all $M$ individuals. Moreover, even in cases where $\grave{\boldsymbol{B}}(r)$ fails to allow full recovery\footnote{Full recovery refers to reconstructing the global model using all local models, whereas partial recovery updates the global model using only a subset (not all) of the local models.}, it remains possible that $\hat{\boldsymbol{B}}(r)$ can successfully decode all individuals, owing to the additional degrees of freedom introduced by $\boldsymbol{\mathcal{T}}_{i_r}(r)$. Overall, it hold that
$\hat{P}\geq\hat{P}_{\mathrm{full}}\geq\hat{P}_M\geq\grave{P}_M$. 
One the other hand, as $\check{\boldsymbol{B}}(r)$ is extracted from  $\grave{\boldsymbol{B}}(r)$, by Lemma \ref{lemma:vertical concentartion}, it holds that 
\begin{align*}
    \forall \boldsymbol{\tau}_{i_r}(r):  \mathrm{rank}(\grave{\boldsymbol{B}}(r))\geq\mathrm{rank}(\check{\boldsymbol{B}}(r)).
\end{align*} 
Consequently, $\grave{P}_M\geq\check{P}_M$. 
To conclude, it holds that 
\begin{align}
    \hat{P}\geq\hat{P}_M\geq\grave{P}_M\geq \check{P}_M.
\end{align}
}
Next, we show that full recovery of the local models is the dominant outcome in GC$^+$. That is, $\hat{P}_{\mathrm{full}}$ dominates among all possibilities when GC$^+$ applies. Let us start by looking at the smallest physical value $\check{P}_M$, which is given by 
\begin{align*}
    \check{P}_M= \sum_{v_r=M}^{(M-s)t_r} {(M-s)t_r\choose v_r} p^{(M-s)t_r-v_r}(1-p)^{v_r}.
    \numberthis
    \label{eq:check_PM}
\end{align*}
By the properties of the binomial distribution, when $p$ is not close to $0$ or $1$, and $(M-s)t_r\gg M$, $\check{P}_M$ is significant compared to $0$ and can approach $1$. Consequently, $\hat{P}_M$ is also significant and can approach $1$. As $\hat{P}_{\mathrm{full}}\geq \hat{P}_M$, so $\hat{P}_{\mathrm{full}}$ dominates in all cases. Hence, we have the following lemma.   
\begin{lemma}[{Dominance of Full Recovery in GC$^+$}]
{If $(M-s)t_r\gg M$, GC$^+$ is able to compute the global model based on \textbf{all} local model updates with high probability close to $1$.}
\end{lemma}
{Note that (\ref{eq:check_PM}), as the lower bound of $\hat{P}_{\mathrm{full}}$ is useful for deriving the following lemma, however, the condition $(M - s)t_r \gg M$ is overly strict. Even when this condition is not strictly satisfied, GC$^+$ still demonstrates strong capability in full recovery of the global model, as shown in Fig. \ref{fig: count for full recovery}. } 
\subsection{Convergence Analysis}
Building on the analysis of the success bound, we establish the following properties of GC$^+$. These results address the intertwined decoding of individual models arising from the decoding mechanism. Details can be found in Appendix B.
\begin{lemma}[Bounded Averaging in Coded Networks]\label{lemma: GC$^+$}
For any algorithm, if the PS can observe all possible sets of the given cardinality with equal probability, 
the update rule of the global model in (\ref{eq: GullGC_PSagg}) is unbiased, i.e., 
\begin{align*}   
\mathbb{E}_{\mathcal{K}_4(r)}\left[\sum_{m\in \mathcal{K}_4(r)} \frac{1}{\lvert \mathcal{K}_4(r)\rvert}  \boldsymbol{g}_{m,r}\Bigg \vert \mathcal{K}_4(r)\neq \emptyset\right]=\sum_{m=1}^M \frac{1}{M}\boldsymbol{g}_{m,r}.
\numberthis
\label{eq:lemma3-1}
\end{align*} 
Additionally, it also holds that 
\begin{align*}   
\mathbb{E}_{\mathcal{K}_4(r)}\left[\sum_{m\in \mathcal{K}_4(r)} \frac{1}{\lvert \mathcal{K}_4(r)\rvert^2} \boldsymbol{g}_{m,r}\Bigg \vert \mathcal{K}_4(r)\neq \emptyset\right]\triangleq\sum_{m=1}^M \Bar{\alpha}_m \boldsymbol{g}_{m,r},
\numberthis
\label{eq:lemma3-2}
\end{align*}
where $\Bar{\alpha}_m=\frac{1}{M\Bar{K}}$. Furthermore, by adopting GC$^+$, $\frac{1}{\Bar{K}}$ is upper bounded by $\frac{1}{\Bar{K}_r}\leq  \frac{\check{P}_M \sum_{m=1}^{M-1}\frac{1}{m}}{1-\min\{ P_O^{t_r}, 1- \check{P}_M \}} +\frac{1}{M} \triangleq \frac{1}{K^\star}$ with $\check{P}_M $ given in (\ref{eq:check_PM}).
\end{lemma}
With the help of Lemma \ref{lemma: GC$^+$} and under Assumptions 1–3, we derive the non-convex convergence rate of GC$^+$.
\begin{theorem}\label{theo:GC$^+$}
Let assumption 1$\sim$3 hold, and choose $\eta=K^\star/(8LTI)^{\frac{1}{2}}$ with $I\leq (TI)^{\frac{1}{4}}/K^{\star\frac{3}{4}}$. By adopting GC$^+$ in CoGC, the optimality gap satisfies
\begin{align*}   
&\frac{1}{T}\sum_{r=1}^T\mathbb{E}\left[\lVert\nabla F(\boldsymbol{g}_r) \rVert^2  \big\vert \mathcal{K}_4(r)\neq \emptyset \right] \leq \\
&\frac{496 L}{11(TIK^{\star})^{\frac{1}{2}}}\left( \mathbb{E}\left[F(\boldsymbol{g}_0)\right]-F^\star \right)\\
&+\frac{31}{88(TI)^\frac{3}{2}K^{\star\frac{1}{2}}}\sum_{r=1}^{T}\sum_{m=1}^{M} \frac{1}{M} J_{m,r}^2\\
&+\left(\frac{39}{88(TIK^{\star})^{\frac{1}{2}}}+\frac{1}{88(TIK^{\star})^{\frac{3}{4}}}\right)\frac{\sigma^2}{b}+\\
&\left( \frac{4}{11(TIK^{\star})^{\frac{1}{2}}}+\frac{1}{22(TIK^{\star})^{\frac{3}{4}}} +\frac{31}{22(TI)^\frac{1}{4}K^{\star\frac{5}{4}}} \right)\hspace{-1mm}\sum_{m=1}^{M}\frac{1}{M} D_m^2.
\numberthis \label{eq:theorem 1}
\end{align*}
\end{theorem}
\begin{proof}[Proof Sketch]
    It can be verified that Lemmas 2–4 from the supplementary material of \cite{10802992} remain valid. Therefore, Theorem~\ref{theo:GC$^+$} can be proved by following similar steps to those of \cite[Theorem 1]{10802992}. 
\end{proof}


\section{Simulations}\label{sec: simulation} 
We run experiments on the MNIST and CIFAR-10, and compare the performance of the following methods in the presence of data heterogeneity. 
\begin{enumerate}[label=(\roman*)]
\item CoGC, as described in Section  \ref{sec: CoGC}. 
\item GC$^+$, as described in Section  \ref{sec: GC$^+$}.
\item FL with perfect connectivity. This benchmark provides insights into the \textbf{ideal} performance the FL system can achieve under heterogeneous data distribution. 
\item FL with intermittent links from clients to PS. The network can be heterogeneous. The update rule follows (\ref{eq: GullGC_PSagg}). 
\end{enumerate} 
To simulate heterogeneous data distributions across clients, we employ two distinct strategies:
\begin{itemize}
    \item MNIST: Each client is assigned data from a single class, resulting in a highly imbalanced and non-IID distribution.
    \item CIFAR-10: We use a Dirichlet distribution-based sampling method to control the level of data imbalance across clients. The concentration parameter $\gamma$ is set to $0.35$, yielding a moderately non-IID distribution.
\end{itemize}
In the simulation, the number of clients is set to $M=10$. Each client is assigned an equal number of data samples. The total number of training rounds is set to $T=100$, and each client performs $I=5$ local training iterations per round.
The coding parameter $s$ is set to $7$ without specification.
Additional hyperparameters and neural network architectures are detailed in Table~\ref{tbl: cnn structures}.
We report the average results of multiple runs. For a fair comparison, the performance of CoGC is truncated at the $100^\mathrm{th}$ training round. 

\begin{table}[!ht]
\caption{
\footnotesize
{NN architecture, hyperparameter specifications}}
\label{tbl: cnn structures}
\resizebox{\linewidth}{!}{
\begin{tabular}{ccc}
\toprule
{\bf Datasets}& 
{\bf MNIST} & 
{\bf CIFAR-10}\\
\toprule
Neural network &
CNN &
CNN \\
Model architecture$^*$ & 
\begin{tabular}{p{.18\textwidth}}
\centering
{\bf C}(1,10)
-- {\bf C}(10,20)
-- {\bf D}
-- {\bf L}(50)
-- {\bf L}(10)
\end{tabular}
&
\begin{tabular}{p{.18\textwidth}}
\centering
{\bf C}(3,32)
-- {\bf R}
-- {\bf M}
-- {\bf C}(32,32)
-- {\bf R}
-- {\bf M}
-- {\bf L}(256)
-- {\bf R}
-- {\bf L}(64)
-- {\bf R}
-- {\bf L}(10)
\end{tabular}
\\
\midrule
Loss function &
\multicolumn{2}{c}{Negative log-likelihood loss (NLLL)}\\
\addlinespace[1ex]
Learning rate $\eta$
&
$\eta=0.005$ & $\eta=0.02$ \\ 
\addlinespace[1ex] 
\midrule
Batch size  &
\multicolumn{2}{c}{1024}\\
\addlinespace[1ex]
\bottomrule
\end{tabular}}
\vskip.2\baselineskip
\begin{tabular}{p{.47\textwidth}}
$^*$
\begin{footnotesize}    
{\bf C}(\# in-channel, \# out-channel): a 2D convolution layer (kernel size 3, stride 1, padding 1);
{\bf R}: ReLU activation function;
{\bf M}: a 2D max-pool layer (kernel size 2, stride 2);
{\bf L}: (\# outputs): a fully-connected linear layer;
{\bf D}: a dropout layer (probability 0.2).
\end{footnotesize}
\end{tabular}
\vspace{-15pt}
\end{table}
\begin{figure*}[t]
    \centering
    \begin{subfigure}[b]{0.32\textwidth}
        \centering
        \includegraphics[width=\linewidth]{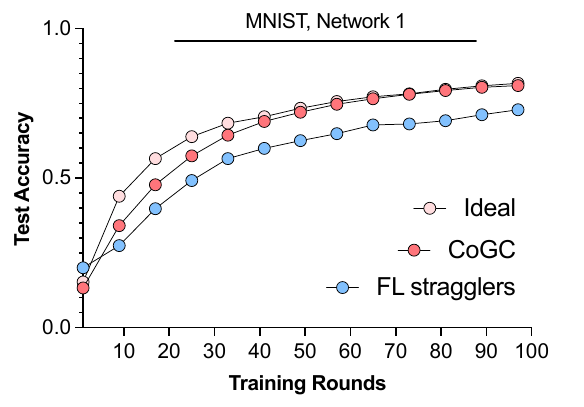}
        \caption{}
        \label{fig:mnist_n1}
    \end{subfigure}
    \hfill
    \begin{subfigure}[b]{0.32\textwidth}
        \centering
        \includegraphics[width=\linewidth]{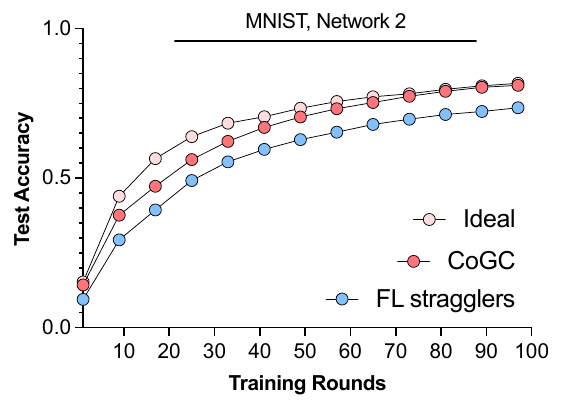}
        \caption{}
        \label{fig:mnist_n2}
    \end{subfigure}
    \hfill
    \begin{subfigure}[b]{0.32\textwidth}
        \centering
        \includegraphics[width=\linewidth]{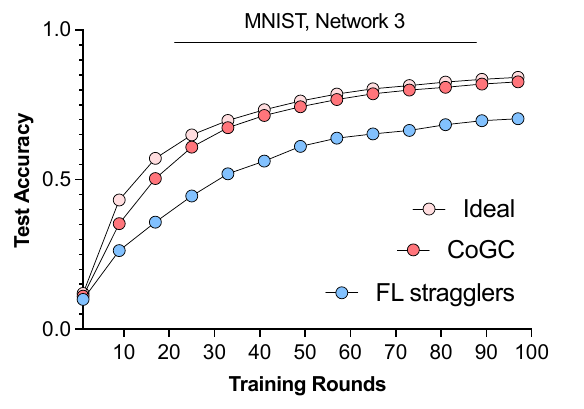}
        \caption{}
        \label{fig:mnist_n3}
    \end{subfigure}
    \caption{{Comparisons of the ideal FL with perfect connectivity, CoGC, and FL with intermittent links on MNIST in different networks.}}
    \label{fig:mnist_all}
    \vspace{-0mm}
\end{figure*}
\begin{figure*}[t]
    \centering
    \begin{subfigure}[b]{0.32\textwidth}
        \centering
        \includegraphics[width=\linewidth]{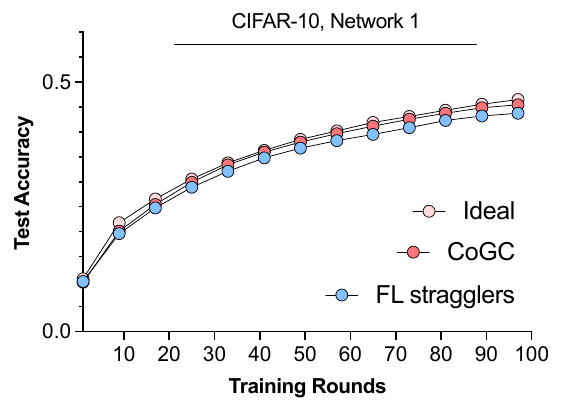}
        \caption{}
        \label{fig:cifar_n1}
    \end{subfigure}
    \hfill
    \begin{subfigure}[b]{0.32\textwidth}
        \centering
        \includegraphics[width=\linewidth]{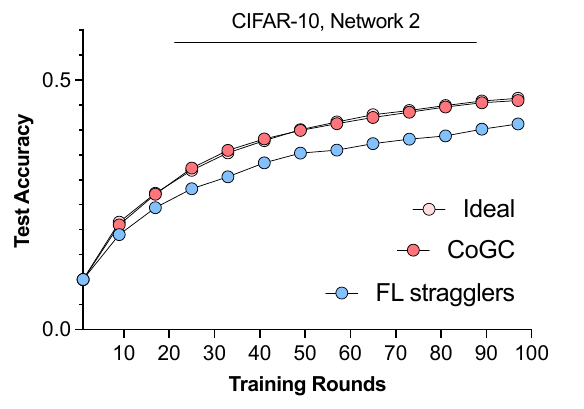}
        \caption{}
        \label{fig:cifar_n2}
    \end{subfigure}
    \hfill
    \begin{subfigure}[b]{0.32\textwidth}
        \centering
        \includegraphics[width=\linewidth]{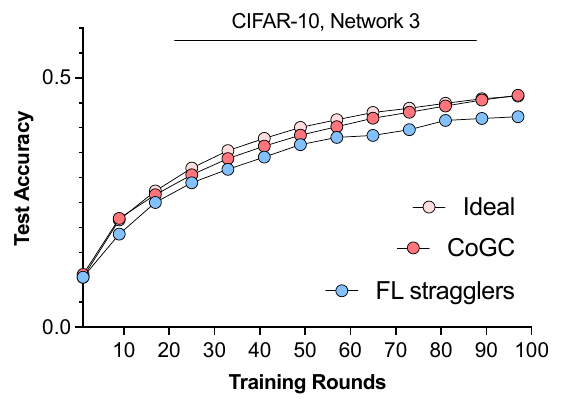}
        \caption{}
        \label{fig:cifar_n3}
    \end{subfigure}
    \caption{{Comparison of the ideal FL with perfect connectivity, CoGC, and FL with intermittent links on CIFAR-10 in different networks.}}
    \label{fig:cifar_all}
\end{figure*}
\begin{figure*}[htbp]
    \centering
       \begin{minipage}[b]{0.49\textwidth}
        \centering    \includegraphics[width=0.7\linewidth]{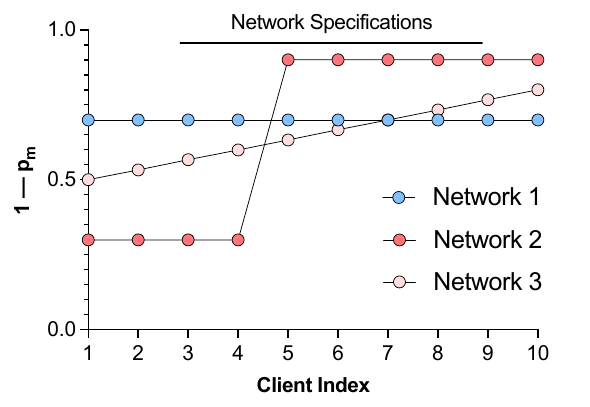}
        \caption{{Specifications of the Networks. When applying to CoGC, $p_{mk}=0.1$, $p_{mk}\sim\mathcal{U}(0.8,1)$, and $p_{mk}\sim\mathcal{U}(0.8,1)$ in Network 1, 2, 3, respectively. }} 
        \label{fig:networks}
    \end{minipage}
    \hfill
    \begin{minipage}[b]{0.49\textwidth}
    \centering
    \includegraphics[width=0.7\linewidth]{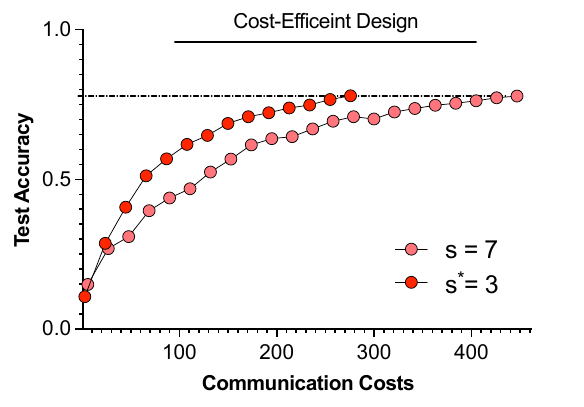}    
    \caption{{Communication cost comparison between the communication-efficient GC and regular GC to achieve the same test accuracy level.}} 
    \label{fig:TestAcc_TradeOff}
    \end{minipage}
    \vspace{-5mm}
\end{figure*}

\subsection{Optimality of CoGC over General Networks}
The performance comparison, measured by test accuracy, between the proposed CoGC and other benchmarks on MNIST and CIFAR-10 is presented in Fig. \ref{fig:mnist_all} and Fig. \ref{fig:cifar_all}, respectively.  The experiments are designed in both homogeneous and heterogeneous networks. The networks are detailed in Fig.\ref{fig:networks}. 
\begin{figure*}
    \centering
    \begin{subfigure}[b]{0.32\textwidth}
        \centering
        \includegraphics[width=\linewidth]{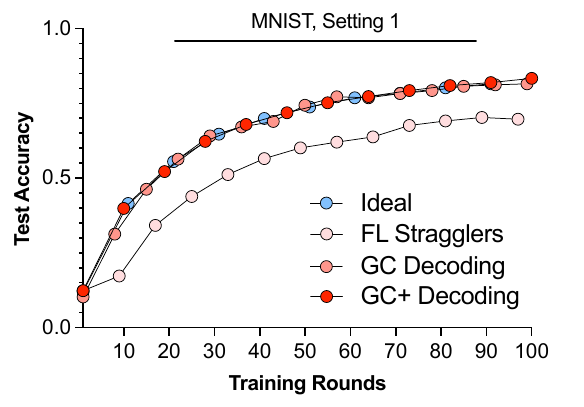}
        \caption{$p_{mk}=0.1$, $p_m=0.4$}
        \label{fig:mnist_gc+1}
    \end{subfigure}
    \hfill
    \begin{subfigure}[b]{0.32\textwidth}
        \centering
        \includegraphics[width=\linewidth]{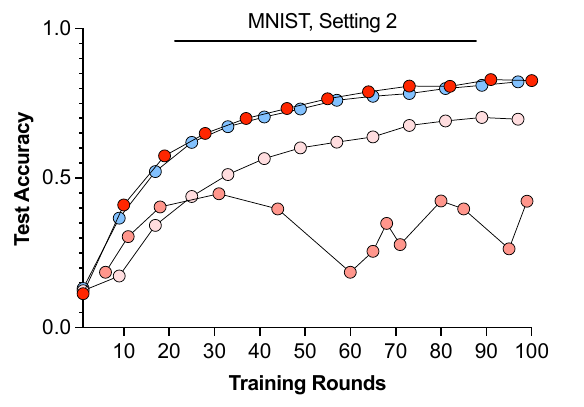}
        \caption{$p_{mk}=0.25$, $p_m=0.4$}
        \label{fig:mnist_gc+2}
    \end{subfigure}
    \hfill
    \begin{subfigure}[b]{0.32\textwidth}
        \centering
        \includegraphics[width=\linewidth]{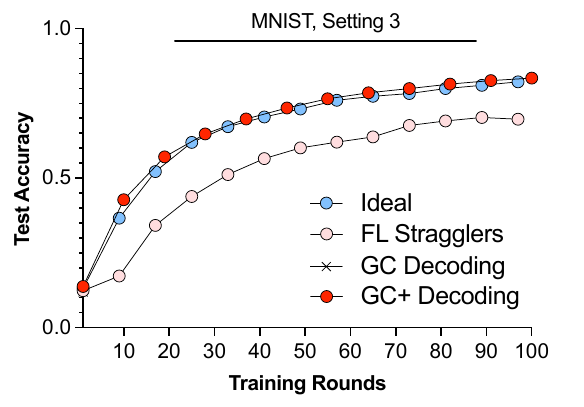}
        \caption{$p_{mk}=0.5$, $p_m=0.4$}
        \label{fig:mnist_gc+3}
    \end{subfigure}
    \caption{{Comparison of ideal FL with perfect connectivity, standard GC decoding, GC$^+$ decoding, and FL with intermittent links on the MNIST dataset, evaluated under three levels of client-to-client connectivity: good (left), moderate (middle), and poor (right). A cross ($\times$) indicates failure to operate. }}
    \label{fig:mnist_gc+}
    \vspace{-3mm}
\end{figure*}
\begin{figure*}
    \centering
    \begin{subfigure}[b]{0.32\textwidth}
        \centering
        \includegraphics[width=\linewidth]{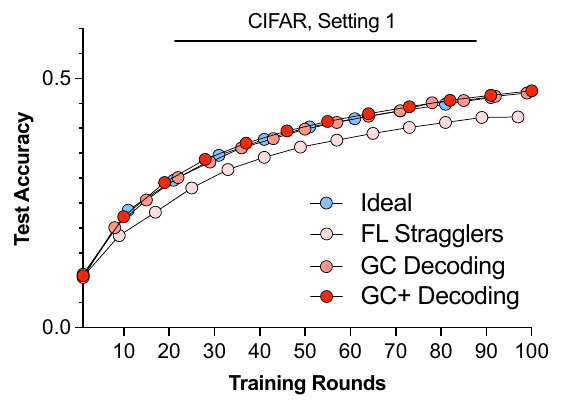}
        \caption{$p_{mk}=0.1$, $p_m=0.4$}
        \label{fig:cifar_gc+1}
    \end{subfigure}
    \hfill
    \begin{subfigure}[b]{0.32\textwidth}
        \centering
        \includegraphics[width=\linewidth]{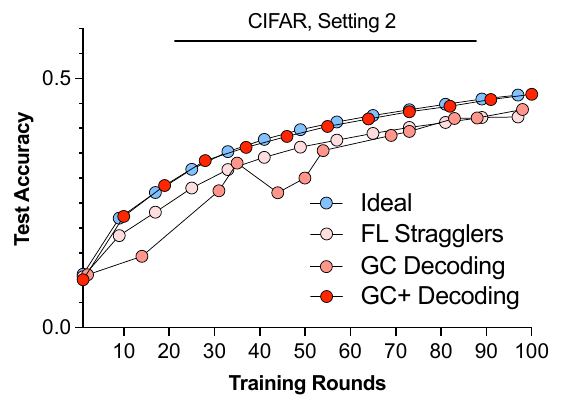}
        \caption{$p_{mk}=0.25$, $p_m=0.4$}
        \label{fig:cifar_gc+2}
    \end{subfigure}
    \hfill
    \begin{subfigure}[b]{0.32\textwidth}
        \centering
        \includegraphics[width=\linewidth]{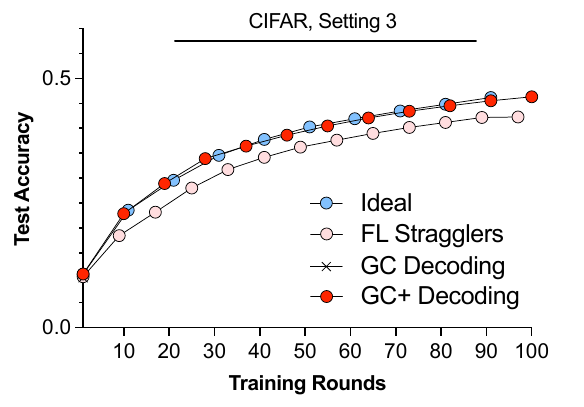}
        \caption{$p_{mk}=0.5$, $p_m=0.4$}
        \label{fig:cifar_gc+3}
    \end{subfigure}
    \caption{{Comparison of ideal FL with perfect connectivity, standard GC decoding, GC$^+$ decoding, and FL with intermittent links on the CIFAR-10 dataset, evaluated under three levels of client-to-client connectivity: good (left), moderate (middle), and poor (right). A cross ($\times$) indicates failure to operate}}
    \label{fig:cifar_gc+}
    \vspace{-3mm}
\end{figure*}
In all settings, CoGC consistently outperforms other methods and reaches the optimal performance defined by perfect connectivity. Additionally, we observe that the convergence speed of CoGC is slightly slower than the ideal case. This observation aligns with our theoretical analysis, which suggests that the convergence rate is reduced by the factor $1-P_O$ \cite{weng2024cooperative}. This is due to the accumulated divergence among consecutively trained local updates. 
Although homogeneous networks do not necessarily lead to sub-optimality, the convergence speed can be significantly reduced by the client drift and the communication opportunities, as shown in Fig. \ref{fig:mnist_n1} and Fig. \ref{fig:cifar_n1}. In contrast, the proposed CoGC prevents the length of computation from being reduced statistically by communications \cite{weng2025heterogeneityawareclientsamplingunified} and ensures the participation of all clients in the aggregation, which reduces the bias of aggregation to some extent. Hence, CoGC effectively mitigates the negative effects caused by unreliable communication. 
In heterogeneous networks (Fig. \ref{fig:mnist_n2}, Fig. \ref{fig:mnist_n3}, Fig. \ref{fig:cifar_n2}, and Fig. \ref{fig:cifar_n3}), FL experiences both slower convergence and inherent sub-optimality, as the global model tends to converge to a biased stationary point instead of the global optimum. In addition to its previously discussed advantages, CoGC further enhances robustness by incorporating a coding structure that guarantees correct aggregation. This mechanism allows CoGC to effectively mitigate the inconsistency caused by communication heterogeneity. As a result, CoGC consistently achieves optimal performance across varying levels of data heterogeneity and various networks.

\vspace{-0.5em}
\subsection{Verification of the Cost-Efficient Design}
The network is configured with $p_m=p_{mk}=0.1$. The target outage probability is set to $P_O^*=0.5$, and the corresponding value of $s^*$ is computed accordingly. Communication costs are measured as the total number of transmissions, as described in Section~\ref{sec:num_comm_cost}.
With a learning rate of $\eta=0.005$,  training is terminated once the test accuracy first reaches $78\%$. As shown in Fig.~\ref{fig:TestAcc_TradeOff}, the proposed cost-efficient design of GC achieves this target with a substantially reduced communication cost by 39.6\%, demonstrating its effectiveness.

\vspace{-0.5em}
\subsection{The Enhanced Reliability of GC$^+$}
{Fig. \ref{fig:Overall_outage} demonstrates that GC decoding exhibits strong resilience to poor client-to-PS connectivity. However, it remains highly sensitive to degraded client-to-PS conditions, as indicated by the consistently high outage probability $P_O$ approaching $1$ across all values of $s$. As discussed in Remark \ref{remark:commu. attempts}, such behavior can hinder the convergence of FL algorithms. This section investigates this effect in detail and evaluates the performance improvements introduced by GC$^+$.
}

{In the simulation, the client-to-PS connectivity is configured to be poor, while the connectivity conditions among clients are varied. The results are presented in Fig. \ref{fig:mnist_gc+} and Fig. \ref{fig:cifar_gc+}. The $t_r$ is set to $2$, for fairness, the number of communication attempts for each GC decoding is also set to $2$. 
With good connectivity among clients, both GC and GC$^+$ perform well and closely approach the ideal baseline. However, under degraded client-to-client communication, the performance of GC decoding deteriorates significantly. In such cases, it may perform worse than standard FL under unreliable communication and even fail to function effectively, as the PS only successfully computes the global model a few times out of $100$ communication rounds. 
This phenomenon aligns with our analysis in Remark \ref{remark:commu. attempts}. Additionally, we observe that under poor client-to-client connectivity, the performance degradation of GC decoding becomes more pronounced as the level of data heterogeneity increases. This phenomenon is consistent with the analysis in Remark~\ref{remark: convergence wrt data hetero}, where it is noted that accumulated divergence becomes more significant in the presence of heterogeneous data.
In contrast, GC$^+$ consistently approaches the ideal performance across all settings. This robustness stems from its ability to benefit from disrupted links, as client-to-client outages can enhance the rank of the received coefficient matrix (as analyzed in Section~\ref{GC$^+$: Outage analysis}).
}
\vspace{-6pt}

\section{Conclusion}\label{sec: conclusion}    
This work studies CoGC, a novel gradient-sharing-based GC framework that eliminates the need for dataset replication and is thus communication-computation-efficient. It does not require prior knowledge about network statistics.
This work presents thorough analyses of the CoGC framework, including a general outage analysis, the convergence bound with $99.86\%$ guarantee, and an assessment of its secure aggregation property. Subsequently, a cost-efficient design problem of the GC matrices is formulated. Despite offering many advantages, the standard GC decoding mechanism may cause resource waste and hinder convergence. To address this, a complementary decoding mechanism is proposed, named GC$^+$, which further improves the straggler tolerance of GC matrices. The complete theoretical analysis of GC$^+$ is performed, including the impact of outages on the GC matrices and the convergence analysis. Finally, all the phenomena of interest and the efficacy of the proposed method are validated through simulations.


\appendices
\vspace{-0.5em}
\section{Proof of Theorem \ref{theo:convergence CoGC}}\label{appx: Proof of Theorem 1}
With the help of Lemma 2$\sim$4 in \cite{weng2024supplementaryfilecooperativegradient}, the first half proof of Theorem 1 is the same as in \cite{weng2024supplementaryfilecooperativegradient}, here, we begin with the preliminary bound acquired in (28) in \cite{weng2024supplementaryfilecooperativegradient}, i.e.,  
\begin{align*}
    &\frac{1}{T}\sum_{r\in [T]} H_1\mathbb{E}\left[\left\lVert\nabla F(\boldsymbol{g}_{r}^{0}) \right\rVert^2\right]\leq \frac{1}{T}H_2\left(F^\star-F(\boldsymbol{g}_{r}^0)\right)\\
    &\hspace{1.5cm}+\frac{1}{T}\sum_{r\in [T]}H_3\sum_{m=1}^M p_m D_m^2
    +\frac{1}{T}\sum_{r\in [T]}H_4 \sigma^2,
\numberthis
\label{eq:theorem1_tau_pre}
\end{align*}
where
\begin{subequations}
\begin{align}
    &H_1=\frac{1}{2}R_r-H_3,\\
    &H_2=\frac{1}{\eta I },\\
    &H_3=2\eta R_r^2 I L+\eta^2\frac{\frac{2}{3}L^2R_r(R_r I\hspace{-1mm}+\hspace{-1mm}1)(2R_r I\hspace{-1mm}+\hspace{-1mm}1)(\frac{1}{2}\hspace{-1mm}+\hspace{-1mm}\eta R_r I L)}{1-R_r I(R_r I+1)\eta^2L^2},\\
    &H_4=\frac{1}{2}\eta R_r L \sum_{m=1}^M p_m^2+\eta^2\frac{\frac{1}{2}L^2R_r(R_r I+1)(\frac{1}{2}+\eta R_r I L)}{1-R_r I(R_r I+1)\eta^2L^2}.
\end{align}
\end{subequations}
By choosing $\eta=\frac{1}{L}\sqrt{\frac{M}{T}}$ and $T$ large, $H_1 \sim H_4$ is dominated by $\mathcal{O}(1/\sqrt{T})$, (\ref{eq:theorem1_tau_pre}) can be approximated as follows, 
\begin{align*}
    \Bar{\mathcal{J}}_1(R_r)\cdot  \min_{r\in [T]} \mathbb{E}\left[\left\lVert\nabla F(\boldsymbol{g}_{r}^{0}) \right\rVert^2\right] \leq \Bar{\mathcal{J}}_2(R_r),
    \label{eq:approx}
    \numberthis
\end{align*}
where
\begin{subequations}
\begin{align*}
    &\Bar{\mathcal{J}}_1(R_r)=\frac{1}{T}\sum_{r\in [T]}\mathcal{J}_1(R_r)=\frac{1}{T}\sum_{r\in [T]} \left( \frac{1}{2}R_r-2I\sqrt{\frac{M}{T}} R_r^2 \right),\numberthis\\
    &\Bar{\mathcal{J}}_2(R_r)= \frac{L}{TI}\sqrt{\frac{T}{M}}\left(F^\star-F(\boldsymbol{g}_{r}^0)\right)+\\
    &
    \hspace{3mm}\frac{1}{T}\sum_{r\in [T]} \underbrace{\left(2I\sqrt{\frac{M}{T}}\sum_{m=1}^M p_m D_m^2 R_r^2 + \frac{\sigma^2}{2} \sqrt{\frac{M}{T}}\sum_{m=1}^{M} p_m^2 R_r\right)}_{\mathcal{J}_3(R_r)}.\numberthis
\end{align*}
\end{subequations}

According to central limit theorem (CLT), when $T$ is chosen large but finite, $\mathcal{J}_1(R_r)$ and $\mathcal{J}_2(R_r)$ are asymptotically Gaussian r.v.s, below we compute the parameters characterizing the Gaussian distributions of $\mathcal{J}_1(R_r)$ and $\mathcal{J}_2(R_r)$, respectively. 
\begin{subequations}
\begin{align*}
    &\mu_{\mathcal{J}_1}=\mathbb{E}\left[ \mathcal{J}_1(R_r) \right]\\
    &\hspace{6mm}=\frac{1-P_O}{P_O}\left(\frac{1}{2}\mathrm{Li}_{-1}(P_O)-2I\sqrt{\frac{M}{T}}\mathrm{Li}_{-2}(P_O) \right),\numberthis \label{eq:E_J1}\\
    &\mathbb{E}\left[ \left(\mathcal{J}_1(R_r)\right)^2 \right]\hspace{-1mm}=\hspace{-1mm}\frac{1-P_O}{P_O}\Bigg(\frac{1}{4}\mathrm{Li}_{-2}(P_O)-2I\sqrt{\frac{M}{T}}\mathrm{Li}_{-3}(P_O)\\
    &\hspace{40mm}+4I^2\frac{M}{T}\mathrm{Li}_{-4}(P_O) \Bigg), \numberthis
    \label{eq:E_J1^2}
\end{align*}
\end{subequations}
where $\mathrm{Li}_{-v}(z)=\sum_{k=1}^\infty k^v z^k$ is the polylogarithm function and specific values can be computed by $\mathrm{Li}_{-v}(z)= \left(z\frac{\partial }{\partial z}\right)^v \frac{z}{1-z} $. In (\ref{eq:E_J1}) and (\ref{eq:E_J1^2}), we use the fact that $\sum_{k=1}^\infty k^v z^{k-1} (1-z)=\frac{1-z}{z} \mathrm{Li}_{-v}(z)$.

Subsequently, the variance $\sigma_{\mathcal{J}_1}^2$ of $\mathcal{J}_1(R_r)$ is given by
\begin{align*}
\sigma_{\mathcal{J}_1}^2=\mathbb{E}\left[\left(\mathcal{J}_1(R_r)\right)^2 \right]-\left(\mathbb{E}\left[ \mathcal{J}_1(R_r) \right]\right)^2.
    \numberthis
    \label{eq:sigma1^2}
\end{align*}
Similarly, we have
\begin{subequations}
\begin{align*}
    &\mu_{\mathcal{J}_3}=\mathbb{E}\left[\mathcal{J}_3(R_r)\right]\\
    &\hspace{6mm}=\frac{1-P_O}{P_O}\Bigg(\frac{\sigma^2}{2}\sqrt{\frac{M}{T}}\sum_{m=1}^Mp_m^2\mathrm{Li}_{-1}(P_O) \\
    &\hspace{2.5cm}+2I\sqrt{\frac{M}{T}}\sum_{m=1}^{M}p_mD_m^2\mathrm{Li}_{-2}(P_O) \Bigg), \label{eq:E_J3}
    \numberthis\\
    &\mathbb{E}\left[ \left(\mathcal{J}_3(R_r)\right)^2 \right]=\frac{1-P_O}{P_O}\Bigg( \frac{M\sigma^4}{4T}\left( \sum_{m=1}^Mp_m^2 \right)^2 \mathrm{Li}_{-2}(P_O)\\
    &\hspace{26mm}+\frac{4MI}{T}\left( \sum_{m=1}^Mp_m D_m^2\right)^2\mathrm{Li}_{-4}(P_O)\\
    &\hspace{5mm}+\frac{2MI}{T}\left( \sum_{m=1}^Mp_m^2 \right)\left( \sum_{m=1}^Mp_m D_m^2\right)\mathrm{Li}_{-3}(P_O)   \Bigg). \label{eq:E_J3^2} \numberthis
\end{align*}
\end{subequations}
Consequently, we have
\begin{subequations}
    \begin{align}
        &\mu_{\mathcal{J}_2}=\mathbb{E}\left[\mathcal{J}_2(R_r)\right]=\frac{L}{TI}\sqrt{\frac{T}{M}}\left(F^\star-F(\boldsymbol{g}_{r}^0)\right)+\mathbb{E}\left[\mathcal{J}_3(R_r)\right], \label{eq:E_J2}\\
        &\sigma_{\mathcal{J}_2}^2=\sigma_{\mathcal{J}_3}^2=\mathbb{E}\left[\left(\mathcal{J}_3(R_r)\right)^2 \right]-\left(\mathbb{E}\left[ \mathcal{J}_3(R_r) \right]\right)^2 \label{eq:sigma2^2}.
    \end{align}
\end{subequations}
According to CLT, we have
\begin{subequations}
\begin{align}
    &\Bar{\mathcal{J}}_1(R_r)\sim \mathcal{N}\left(\mu_{\mathcal{J}_1}, \frac{\sigma_{\mathcal{J}_1}^2}{T} \right),\\
    &\Bar{\mathcal{J}}_2(R_r)\sim \mathcal{N}\left(\mu_{\mathcal{J}_2}, \frac{\sigma_{\mathcal{J}_2}^2}{T}\right)
\end{align}    
\end{subequations}
where $\mu_{\mathcal{J}_1}$. $\sigma_{\mathcal{J}_1}^2$, $\mu_{\mathcal{J}_2}$, $\sigma_{\mathcal{J}_2}^2$ are given in (\ref{eq:E_J1})$\sim$(\ref{eq:sigma2^2}). 

A simple variation of (\ref{eq:approx}) gives
\begin{align*}
    \min_{r\in [T]} \mathbb{E}\left[\left\lVert\nabla F(\boldsymbol{g}_{r}^{0}) \right\rVert^2\right] \leq \frac{\Bar{\mathcal{J}}_1(R_r)}{\Bar{\mathcal{J}}_2(R_r)}.
    \numberthis
\end{align*}
Having determined the distribution of $\Bar{\mathcal{J}}_1(R_r)$ and $\Bar{\mathcal{J}}_2(R_r)$, with the help of Delta method, one can further approximate the distribution of $\frac{\Bar{\mathcal{J}}_2(R_r)}{\Bar{\mathcal{J}}_2(R_r)}$. First, approximate the non-linear ratio $\frac{\Bar{\mathcal{J}}_2(R_r)}{\Bar{\mathcal{J}}_1(R_r)}$ by linear functions around $\Bar{\mathcal{J}}_1(R_r)=\mu_{\mathcal{J}_1}, \Bar{\mathcal{J}}_2(R_r)=\mu_{\mathcal{J}_2}$ using first-order Taylor expansion as
\begin{align*}
    &\frac{\Bar{\mathcal{J}}_2(R_r)}{\Bar{\mathcal{J}}_1(R_r)} \approx \frac{\mu_{\mathcal{J}_2}}{\mu_{\mathcal{J}_1}}+\frac{1}{\mu_{\mathcal{J}_1}}\left(\Bar{\mathcal{J}}_2(R_r)- \mu_{\mathcal{J}_2} \right)\\
    &\hspace{2.3cm}-\frac{\mu_{\mathcal{J}_2}}{\mu_{\mathcal{J}_1^2}}\left(\Bar{\mathcal{J}}_1(R_r)- \mu_{\mathcal{J}_1} \right),
    \numberthis
\end{align*}
So $\frac{\Bar{\mathcal{J}}_2(R_r)}{\Bar{\mathcal{J}}_2(R_r)}$ can be approximated as a Gaussian r.v. characterized by 
\begin{subequations}
    \begin{align*}
        &\mathbb{E}\left[ \frac{\Bar{\mathcal{J}}_2(R_r)}{\Bar{\mathcal{J}}_1(R_r)} \right]=\frac{\mu_{\mathcal{J}_2}}{\mu_{\mathcal{J}_1}},\numberthis\\
        &\sigma_{\Bar{\mathcal{J}}_2/\Bar{\mathcal{J}}_1}^2 \approx \frac{\sigma_{\mathcal{J}_2}^2/T}{\mu_{\mathcal{J}_1}^2}+\frac{\mu_{\mathcal{J}_2}^2\sigma_{\mathcal{J}_1}^2/T}{\mu_{\mathcal{J}_1}^4}\\
        &\hspace{26mm}-2\frac{\mu_{\mathcal{J}_2}}{\mu_{\mathcal{J}_1}^3}\mathrm{cov}\left( \Bar{\mathcal{J}}_2(R_r), \Bar{\mathcal{J}}_1(R_r)\right),
        \numberthis
    \end{align*}
\end{subequations}
when $T$ is large. 
It is difficult to give a closed-form expression of $\mathrm{cov}\left( \Bar{\mathcal{J}}_2(R_r), \Bar{\mathcal{J}}_1(R_r)\right)$, but it is possible to upper bound the $\sigma_{\Bar{\mathcal{J}}_2/\Bar{\mathcal{J}}_1}^2$. By Cauchy-Schwarz Inequality, it holds that 
\begin{align}
    \left\lvert\mathrm{cov}\left( \Bar{\mathcal{J}}_2(R_r), \Bar{\mathcal{J}}_1(R_r)\right)\right\rvert \leq \frac{\sigma_{\mathcal{J}_1} \sigma_{\mathcal{J}_2}}{T},
\end{align}
thus,
\begin{align}
    &\sigma_{\Bar{\mathcal{J}}_2/\Bar{\mathcal{J}}_1}^2\leq \frac{\sigma_{\mathcal{J}_2}^2}{\mu_{\mathcal{J}_1}^2T}+\frac{\mu_{\mathcal{J}_2}^2\sigma_{\mathcal{J}_1}^2}{\mu_{\mathcal{J}_1}^4T}+2\frac{\mu_{\mathcal{J}_2}\sigma_{\mathcal{J}_1} \sigma_{\mathcal{J}_2}}{\mu_{\mathcal{J}_1}^3 T}\overset{\triangle}{=} \sigma_{\mathrm{max}}^2.
    \label{eq:upperbound_variance_ratio}
\end{align}
To summarize, 
\begin{align*}
    \frac{\Bar{\mathcal{J}}_2(R_r)}{\Bar{\mathcal{J}}_1(R_r)}\sim \mathcal{N}\left(\frac{\mu_{\mathcal{J}_2}}{\mu_{\mathcal{J}_1}} , \sigma_{\Bar{\mathcal{J}}_2/\Bar{\mathcal{J}}_1}^2\right).
    \numberthis
\end{align*}
According to Three-Sigma Rule (3SR) and (\ref{eq:upperbound_variance_ratio}), $\frac{\Bar{\mathcal{J}}_2(R_r)}{\Bar{\mathcal{J}}_1(R_r)}\leq \frac{\mu_{\mathcal{J}_2}}{\mu_{\mathcal{J}_1}}+3\sigma_{\mathrm{max}}^2 $ w.p. larger than $99.86\%$.


\section{Proof of Lemma \ref{lemma: GC$^+$}}
\begin{proof}[Proof of (\ref{eq:lemma3-1})]
{First, we prove the unbiasedness of the global model recovery. In GC$^+$, when the standard GC decoding mechanism can work, (\ref{eq: global_update_at_PS}) is always met, and it can be categorized into the case when $\mathcal{K}_4(r)=[M]$. If the GC fails, the complementary decoding mechanism applies, $\mathcal{K}_4(r)$ can be any non-empty subset of $[M]$. Let $\mathcal{K}_4(r)$ contain all the decodable local models in $r$-th round and $\mathcal{K}_4^c(r)=[M]\setminus \mathcal{K}_4(r)$ contains all local models that can not be decoded by the PS. The unbiasedness of the updated rule in GC$^+$ is proved as follows,} 
\begin{subequations}
\begin{align}   
&\mathbb{E}_{\mathcal{K}_4(r)}\left[\sum_{m\in \mathcal{K}_4(r)} \frac{1}{\lvert \mathcal{K}_4(r)\rvert}  \boldsymbol{g}_{m,r}\Bigg \vert \mathcal{K}_4(r)\neq \emptyset\right]\\
&=\mathbb{E}_{K_r}\left[ \mathbb{E}_{\substack{ \mathcal{K}_4(r): \\ \lvert\mathcal{K}_4(r)\rvert=K_r }}\left[ \frac{1}{K_r}\sum_{m\in \mathcal{K}_4(r)}  \boldsymbol{g}_{m,r}\right] \right] \label{eq:client sampling}\\
&=\mathbb{E}_{K_r}\left[ \frac{1}{M}\sum_{m=1}^M \boldsymbol{g}_{m,r}\right]=\frac{1}{M}\sum_{m=1}^M \boldsymbol{g}_{m,r}.\label{eq:client sampling unbiased}
\numberthis
\end{align}
\end{subequations}
{For any possible sets $\mathcal{K}_4(r): \lvert\mathcal{K}_4(r)\rvert=K_r\neq 0$, $\mathcal{K}_4(r)$ can be viewed as a specific realization of uniformly sampling $K_r$ clients from $[M]$ without replacement, but induced by unreliable communications in the homogeneous network. Therefore, the inside layer of expectation in (\ref{eq:client sampling}) is unbiased and gives (\ref{eq:client sampling unbiased}), following the conclusion of Scheme II in \cite{li2019convergence}. }  
\end{proof}

\begin{proof}[Proof of (\ref{eq:lemma3-2})]
{Next, we prove the second conclusion in Lemma \ref{lemma: GC$^+$}. This proof partly refers to \cite[Appendix A]{wang2021quantized}, but ours tackles the challenge of the coded networks, where the decoding of individuals is intertwined due to coding blocks.} 
By expanding (\ref{eq:lemma3-2}), it holds that 
\begin{subequations}
\begin{align}
&\mathbb{E}_{\mathcal{K}_4(r)}\left[\sum_{m\in \mathcal{K}_4(r)} \frac{1}{\lvert \mathcal{K}_4(r)\rvert^2} \boldsymbol{g}_{m,r}\Bigg \vert \mathcal{K}_4(r)\neq \emptyset\right] \triangleq\sum_{m=1}^M \Bar{\alpha}_m \boldsymbol{g}_{m,r} \label{eq: 46a}\\
&=\sum_{K_r=1}^{M} \sum_{\substack{ \mathcal{K}_4(r): \\ \lvert\mathcal{K}_4(r)\rvert=K_r }} P(\mathcal{K}_4(r)) \cdot \frac{1}{K_r^2} \cdot\sum_{m\in \mathcal{K}_4(r)} \hspace{-3mm}\boldsymbol{g}_{m,r} \label{eq: 46b}\\
&=\sum_{K_r=1}^{M} P(K_r) \cdot \frac{1}{K_r^2} \cdot \sum_{\substack{ \mathcal{K}_4(r): \\ \lvert\mathcal{K}_4(r)\rvert=K_r }} \sum_{m\in \mathcal{K}_4(r)}\hspace{-3mm}\boldsymbol{g}_{m,r} \label{eq: simple variation}\\
&=\sum_{K_r=1}^{M} P(K_r) \cdot \frac{1}{K_r^2} \cdot \sum_{m=1}^M {M-1 \choose K_r-1}\cdot\boldsymbol{g}_{m,r} \label{eq: change count}\\
&=\sum_{K_r=1}^{M} {M \choose K_r} P(K_r) \cdot \frac{1}{K_r} \cdot \frac{1}{M}\sum_{m=1}^M \boldsymbol{g}_{m,r} \label{eq: com_rule}
\end{align}
\label{eq: alpha}
\end{subequations}
For homogeneous networks, $P(\mathcal{K}_4(r))$ is identical for every possible $\mathcal{K}_4(r): \lvert\mathcal{K}_4(r)\rvert=K_r\neq 0$, and depends only on the cardinality $K_r$. Let $ P(\mathcal{K}_4(r))=P(K_r)$, then we have (\ref{eq: simple variation}). The last two sums $\sum_{\substack{ \mathcal{K}_4(r):  \lvert\mathcal{K}_4(r)\rvert=K_r }} \sum_{m\in \mathcal{K}_4(r)} \boldsymbol{g}_{m,r}$ searches over all possible $\mathcal{K}_4(r)$ and sum up the local models $\boldsymbol{g}_{m,r}$ in all possible $\mathcal{K}_4(r)$. By maintaining the same search space, we can instead count the occurrence frequency of each $\boldsymbol{g}_{m,r}$ in all possible $\mathcal{K}_4(r)$ and sum them up. Hence, we obtain (\ref{eq: change count}). The (\ref{eq: com_rule}) holds due to $\frac{1}{M}{M\choose K_r}=\frac{1}{K_r}{M-1 \choose K_r-1}$.  

The (\ref{eq: alpha}) holds for any choice of $\boldsymbol{g}_{m,r}$. To compute $\Bar{\alpha}_m$, we can set $\boldsymbol{g}_{m,r}=1$. Then (\ref{eq: 46a}) gives
\begin{subequations}
\begin{align}
    &\mathbb{E}_{\mathcal{K}_4(r)}\left[\sum_{m\in \mathcal{K}_4(r)} \frac{1}{\lvert \mathcal{K}_4(r)\rvert^2} \cdot 1 \Bigg \vert \mathcal{K}_4(r)\neq \emptyset\right]\\
    &=\mathbb{E}_{\mathcal{K}_4(r)}\left[ \frac{1}{\lvert \mathcal{K}_4(r)\rvert}  \right]\triangleq \frac{1}{\Bar{K}_r}\label{eq: bar_k_ext}\\
    &=\sum_{K_r=1}^{M} {M \choose K_r} P(K_r) \cdot \frac{1}{K_r} \label{eq: bar_k_exp}
\end{align}
\end{subequations}
By comparing (\ref{eq: 46a}), (\ref{eq: com_rule}), and (\ref{eq: bar_k_ext}), we get
\begin{align}
     \Bar{\alpha}_m=\frac{1}{M\Bar{K}_r}. 
     \label{eq: bar_am}
\end{align}
Define $P_m={M \choose m} P(m)$ as the total probability of decoding $m$ local models. Both the GC decoding and the complementary contribute to $P_M=P(M)$, as the cases where $\tilde{\boldsymbol{B}}_{i_r}$ contains $M-s$ unperturbed rows and where $\tilde{\boldsymbol{B}}$ is of rank $M$ overlap partially. Instead, $P_m$ solely depends on GC$^+$. For $m\in[M]: P_m=\frac{\hat{P}_m}{1-P_{\emptyset}}$,  where $P_{\emptyset}$ denotes the probability of decoding nothing in GC$^+$. The scaling $\frac{1}{1-P_{\emptyset}}$ is due to the conditioning. Moreover, it holds that $P_{\emptyset}\leq \min\{ P_O^{t_r}, 1- \check{P}_M \}$.  
When $(M-s)t_r\gg M$, $\check{P}_M$ dominates $\hat{P}$, as analyzed in Section \ref{sec: Success Bound}. Since that $\hat{P}=\sum_{m=1}^{M} \hat{P}_m$, we have $\forall m\neq M: \hat{P}_m \leq \check{P}_M$. Consequently, $P_m\leq \frac{\check{P}_M}{1-P_{\emptyset}}$. Moreover, $P_M<1$. 
By substituting these relations into (\ref{eq: bar_k_exp}), we have 
\begin{align}
    \frac{1}{\Bar{K}_r}\leq \frac{\check{P}_M \sum_{m=1}^{M-1}\frac{1}{m}}{1-\min\{ P_O^{t_r}, 1- \check{P}_M \}} +\frac{1}{M},
    \label{eq: upper bound Kbar}
\end{align}
where $\check{P}_M$ is given in (\ref{eq:check_PM}). By (\ref{eq: bar_k_ext}), (\ref{eq: bar_am}) and (\ref{eq: upper bound Kbar}), we complete the proof of Lemma \ref{lemma: GC$^+$}. 
\end{proof}

\section{Rank Analysis}\label{appendix: rank_analysis}
The fact that $ \mathrm{rank}(\hat{\boldsymbol{B}}_{i_r})=M-s$ implies that $(i)$ $\forall \mathcal{O}\subset [M]$ where $\lvert \mathcal{O} \rvert=M-s$, there exists $c^{\mathcal{O}, i'_r}_{i_r}\neq 0$, $i_r\in \mathcal{O}$ such that $\forall i'_r\notin\mathcal{O}$, it holds that 
\begin{align*} 
    \boldsymbol{b}_{i'_r}=\sum_{i_r\in \mathcal{O}} c^{\mathcal{O}, i'_r}_{i_r} \boldsymbol{b}_{i_r},
    \numberthis
    \label{eq: linear_independence_B}
\end{align*}
and that $(ii)$ $\forall \mathcal{O}\subset [M]$ where $\lvert \mathcal{O} \rvert>M-s$, $\mathrm{rank}(\mathcal{O})=M-s$. 
If at least $M-s$ rows are unperturbed in $\tilde{\boldsymbol{B}}_{i_r}$, let $\mathcal{U}$ be the set of indices of any $M-s$ unperturbed rows. Then, any perturbed row, say $\tilde{i}_r$-th row, can be expressed as linear combinations of the $M-s$ unperturbed rows such that the unerased entry $b_{\tilde{i}_r, j_r}$ becomes $0$, i.e., 
\begin{align*}
    0=b_{\tilde{i}_r, j_r}-\sum_{i_r\in \mathcal{U}} c^{\mathcal{U}, \tilde{i}_r}_{i_r} b_{i_r, j_r},
    \numberthis
    \label{eq:tranform_unpert}
\end{align*}
and that the $0$ resulting from the erased $b_{\tilde{i}_r,\tilde{j}_r}$ is transformed to $-b_{\tilde{i}_r,\tilde{j}_r}$, i.e., 
\begin{align*}
    -b_{\tilde{i}_r,\tilde{j}_r}=0-\sum_{i_r\in \mathcal{U}} c^{\mathcal{U}, \tilde{i}_r}_{i_r} b_{i_r,\tilde{j}_r}.  
    \numberthis
     \label{eq:tranform_pert}
\end{align*}
Both (\ref{eq:tranform_unpert}) and (\ref{eq:tranform_pert}) is due to  (\ref{eq: linear_independence_B}). Let $\Bar{\boldsymbol{B}}_{i_r}(r)$ denote the matrix after transformation. The rank remains unchanged before and after the transformation, i.e., $\mathrm{rank}(\tilde{\boldsymbol{B}}_{i_r})=\mathrm{rank}(\Bar{\boldsymbol{B}}_{i_r}(r))$. 

The rank of $\Bar{\boldsymbol{B}}_{i_r}(r)$ is counted as follows due to the special structure inherited from $\tilde{\boldsymbol{B}}_{i_r}$. First, the unperturbed part has rank $M-s$.           
Let $J$ denote the number of distinct nonzero patterns of the perturbed rows in $\Bar{\boldsymbol{B}}_{i_r}(r)$. If $n_r^j$ rows correspond to the $j$-th nonzero pattern, each containing $n_c^j$ nonzero elements, the rank contribution of these rows is 
$\min\{ n_r^j, n_c^j, M-s \}$, as (\ref{eq: linear_independence_B}) will hold for rows of identical nonzero pattern if $\min\{ n_r^j, n_c^j,\}>M-s$. Then we remove any $n_r^j-\min\{ n_r^j, n_c^j, M-s \}$ rows of the $j$-th nonzero pattern and keep only $M-s$ unperturbed rows indexed by $\mathcal{U}$, as they do not contribute to the rank. After doing this to all $J$ nonzero patterns, denote the resulting matrix as $\check{\boldsymbol{B}}_{i_r}(r)$. Since the removal preserves rank, $\mathrm{rank}(\check{\boldsymbol{B}}_{i_r}(r))=\mathrm{rank}(\Bar{\boldsymbol{B}}_{i_r}(r))$. The rank of $\Bar{\boldsymbol{B}}_{i_r}(r)$ is given by $n_{i_r}+(M-s)$ due to the full-rank property of random generation \cite{tandon2017gradientcoding}, where $n_{i_r}$ represents the maximum number of the nonzero elements that are neither in the same row nor the same column in the perturbed part.

\balance
\bibliographystyle{IEEEtran}
\bibliography{IEEEabrv,reference}


\end{document}